\documentclass[12pt,reqno,fleqn]{amsart}

\usepackage{array}
\usepackage[mathscr]{euscript}
\usepackage{breqn}

\usepackage{youngtab}
\usepackage{amsmath}
\usepackage{amssymb}
\usepackage{amsxtra}

\usepackage{xcolor}

\usepackage{hyperref}
 \hypersetup{
     colorlinks = true,
    extension = notused,  
     linkcolor = blue,
     anchorcolor = red,
     citecolor = blue,
     filecolor = blue,
     pagecolor = red,
     urlcolor = blue
 }

\usepackage{tikz}
\usepackage{helvet}
\usepackage{listings}
\usepackage{slashed}
\usepackage{cancel}

\usepackage{setspace}
\usepackage{mathtools}

\usepackage[numbers,sort&compress]{natbib}

\newcommand {\la} {\left \langle}
\newcommand {\ra} {\right \rangle}

\newcommand {\CalA} {\mathcal A}

\newcommand {\CalE} {\mathcal E}

\newcommand {\CalM} {\mathcal M}

\newcommand {\BC}   {\mathbb C}

\newcommand {\BR}   {\mathbb R}
\newcommand {\BP}   {\mathbb P}

\newcommand {\BZ}   {\mathbb Z}

\newcommand{\ep}{\epsilon}
\newcommand{\ve}{\varepsilon}

\newcommand{\g}{\mathfrak{g}}
\newcommand{\h}{\mathfrak{h}}

\DeclareMathOperator{\tr} {tr}
\DeclareMathOperator{\Tr} {Tr}

\DeclareMathOperator{\ind} {ind}

\newcommand{\eu}{\mathrm{eu}}

\newcommand{\gym}{g_{\text{\tiny \textsc{ym}}}}

\newcommand{\const}{\mathrm{const}}

\usepackage[margin=1.2in]{geometry}

\begin{document}


\title{Localization for $\mathcal{N}=2$ supersymmetric gauge theories in four
  dimensions}

\author{Vasily Pestun}

\bigskip
\bigskip
\address{IHES, France}
\email{pestun@ihes.fr}

\begin{abstract}
We review the supersymmetric localization of $\mathcal{N}=2$
  theories on curved backgrounds in four dimensions using
  $\mathcal{N}=2$ supergravity and  generalized conformal
Killing spinors. We review some known backgrounds and give examples
of new geometries such as local $T^2$-bundle fibrations. 
We discuss in detail a topological four-sphere
with generic $T^2$-invariant metric.    This review 
is a contribution to the special volume on recent developments in
$\mathcal{N} = 2$ supersymmetric gauge theory and the 2d-4d
relation.
\end{abstract}

\maketitle

\tableofcontents

\numberwithin{equation}{section}
\numberwithin{figure}{section}
\numberwithin{table}{section}

\bibliographystyle{utphys}
\newcommand{  \cgamma} { ^{c}\gamma}
\newcommand{\mycolor}[1]{}
\newcommand{  \conj}[1]{{#1}}
\newcommand{\Cl}{\mathbf{Cl}}
\newcommand{\Mat}{\mathbf{Mat}}
\newcommand{\RS}{\mathrm{R}}

\newcommand{\Gf}{F}
\newcommand{\Tf}{T_{F}}
\newcommand{\Gg}{G}
\newcommand{\GL}{L}
\newcommand{\Tequiv}{\mathsf{T}}
\newcommand{\Gequiv}{\mathsf{G}}

\renewcommand{\gg}{\mathfrak{g}}

\newcommand{\ii}{\mathrm{i}}
\newcommand{\Ver}{I}
\newcommand{\Aut}{\mathrm{Aut}}

\section{Introduction}

Non-perturbative exact results in interacting quantum field theories (QFTs) are
rare and precious and usually we explain them using non-trivial
symmetries of QFT, such as quantum groups in the theory of quantum
integrable systems \cite{Sklyanin:1978pj,Drinfeld:1985,Jimbo:1985}.
 Another instrumental symmetry for exact results in  QFTs is
 supersymmetry. For example, 
Seiberg-Witten solution \cite{Seiberg:1994rs} 
of four-dimensional $\mathcal{N}=2$ supersymmetric field theories is
explained by rigid constraints imposed by $\mathcal{N}=2$ supersymmetry
on the low-energy effective Lagrangian and certain assumptions on electric-magnetic duality. 
For a review of $\mathcal{N}=2$ four-dimensional theory from the modern
angle of view see contributions \cite{Ga} and \cite{N} in this volume.

Exact non-perturbative results in supersymmetric QFTs are suited for strong tests of non-perturbative dualities
between QFTs that have different microscopic description, they give a practical
approximation to interesting physical phenomena in non-supersymmetric
QFTs, and they open new perspectives on fascinating geometrical spaces such moduli spaces of instantons, monopoles,
complex structures, flat connections, and others. 

A fruitful non-perturbative approach is supersymmetric localization. In
finite dimensional
geometry localization appeared in the Lefchetz fixed-point formula, 
Duistermaat-Heckman and Atiyah-Bott formula for integration
of equivariantly closed differential forms \cite{Atiyah-Bott:1984}. In \cite{Witten:1982im}
Witten generalized the localization formula for the infinite-dimensional
geometry of the path integral of supersymmetric quantum
mechanics. Similar approach was proposed to  two-dimensional sigma
models \cite{Witten:1988xj}, four-dimensional gauge theories
\cite{Witten:1988ze} and others. The similarity of these
constructions is the topological twist of a given supersymmetric QFT.
 The topological twist introduces a certain background connection for the
local internal R-symmety of the theory. Usually, this connection is such
that there exists a scalar fermionic supersymmetry generator $Q$ for a
QFT coupled to a generically curved background metric. In
topologically twisted theories the stress-energy tensor is $Q$-exact,
 and, consequently, the theory is metric independent.
A further twist to the supersymmetric  localization of gauge theories, called $\ep$-equivariant deformation or
$\Omega$-background  $\BR^4_{\ep_1, \ep_2}$,
has been added by Nekrasov \cite{Nekrasov:2002qd} based on the
considerations of \cite{Losev:1997tp,Moore:1997dj,Lossev:1997bz,Losev:1995cr}. The
construction of the gauge theory instanton partition function is reviewed
in \cite{Ta} of this volume. 
 The $\ep$-equivariant partition function $Z_{\ep_1,\ep_2}$, referred as
 Nekrasov's function, turned out to be a fascinating object of mathematical
physics, with profound connections to other branches of research such as 
 topological strings (see review \cite{A,KW} in this volume), 
 matrix models (see  review  \cite{M} in this volume), quantum groups
 \cite{Drinfeld:1985,Jimbo:1985}  and integrable systems
 \cite{Faddeev:1987ih}. For a recent study of the instanton
 partition function $Z_{\ep_1, \ep_2}$ for a large class of quiver
 theories see \cite{Nekrasov:2012xe,Nekrasov:2013xda}.  A profound
 connection between four-dimensional gauge theory supersymmetric objects (BPS)
 and two-dimensional conformal field theories (CFT), called BPS/CFT
 correspondence  in \cite{Nekrasov:2004sem},  was a subject of long research
\cite{Nakajima:1994r,Nakajima:1998,Vafa:1994tf,Losev:1995cr,Nekrasov:2002qd,Losev:2003py,Nekrasov:2003rj}.

Another version of localization was used in \cite{Pestun:2007rz} for
 $\mathcal{N}=2$ supersymmetric gauge theory on a
 four-sphere $S^4$ with an insertion of a Wilson operator
 \cite{Pestun:2007rz} or t'Hooft
 operators \cite{Gomis:2011pf}.
 The topological twist is not necessary  because
of rich $OSp(2|4)$ symmetry that  $\mathcal{N}=2$ QFT on $S^4$ has. A similar localization was later
performed for gauge theories on $S^3$ \cite{Kapustin:2009kz},  on $S^2$
\cite{Benini:2012ui,Doroud:2012xw}, on $S^5$
\cite{Kallen:2012va,Kim:2012ava}, on squashed $S^3_b$ \cite{Hama:2011ea,Imamura:2011wg},
on squashed $S^4_{\ep_1,\ep_2}$ \cite{Hama:2012bg} and other geometries. For a review
of 3d localization in this volume see \cite{H}, for a review of  line
operators (such as Wilson and 't Hooft operators) in 4d gauge theory see
review \cite{O} in this volume, and for review of surface operators see
\cite{Gu}. The four-sphere partition function of the $\mathcal{N}=2$ gauge theory of
class $\mathcal{S}_{\mathfrak{g}}$ (see \cite{Ga} in this volume) turned out to be equal to the correlation function of
the 2d conformal $\mathfrak{g}$-Toda theory, the statement known as AGT
conjecture
 \cite{Alday:2009aq}, which provided explicit  beautiful realization
 of the 4d/2d BPS/CFT correspondence.   For a review of AGT conjecture
 (4d/2d BPS/CFT correspondence) in this volume see  \cite{Te}, for review of the
 superconformal index see \cite{RR} and for a review of the 3d/3d version of the
BPS/CFT correspondence see \cite{D}.

A general procedure to construct a QFT on a curved
manifold with some amount of supersymmetry is to couple QFT
with supergravity, choose the supegravity background fields in such a
way that there exists a non-trivial supersymmetric variation under which
these background fields are invariant and then freeze the supegravity
fields. This construction was explored for $\mathcal{N}=1$
supersymmetric four-dimensional theories in \cite{Festuccia:2011fk}. 

In this note we will partially analyze 
off-shell $\mathcal{N}=2$ supersymmetry backgrounds suitable for
localization, and review the case of the four-sphere \cite{Pestun:2007rz}
with a generic $T^2$-invariant  deformation of the metric. We employ the formalism of $\mathcal{N}=2$ supergravity known as
superconformal tensor calculus, see \cite{deWit:1979ug,deWit:1980,
  deWit:1984px,deWit:1983xe} and reviews
\cite{Mohaupt:2000mj,supergravity:2012}. For previous analysis of
$\mathcal{N}=2$ supegravity localization backgrounds see \cite{Gupta:2012cy,Dabholkar:2010uh,Klare:2013dka}.

\subsection*{Acknowledgements}
The author is grateful to Kazuo Hosomichi for the correspondence while
preparing this review and discussion of the results of
\cite{Hama:2012bg}, and Takuya Okuda for the elaborate comments on the
manuscript.  

\section{$\mathcal{N}=2$ supergravity}
\subsection{Gravity multiplet}
 A way to construct  $\mathcal{N}=2$ Poincare supergravity is
to promote  the $\mathcal{N}=2$ superconformal symmetry  to  local
gauge symmery and introduce associated gauge fields.
The gauge fields are combined with auxiliary fields to form
\emph{Weyl} multiplet. The notations are collected in the table \ref{ta:fields}.

\renewcommand{\arraystretch}{1.5}
\begin{table}[!h]
\caption{The \emph{Weyl}  multiplet \label{ta:fields}
}
  \begin{tabular}{lccc}
symmetry & gauge field & constraint & parameter  \\
\hline
  translation  &  $e^{\hat \mu}_\mu$  \\  
 rotation       & $\omega^{\hat \mu \hat \nu }_{\mu}$ &   $\hat R(e^{\hat \mu}_\mu)$ \\
    special conformal  & $f^{\hat \mu}_{\mu}$ &  $ \hat R(\omega^{\hat \mu \hat \nu }_{\mu})$ \\
    dilatation                & $b_{\mu}$\\
      translational supersymmetry $Q$ & $\psi^i_\mu$ &  & $\ve^i$\\
     conformal supersymmetry  $S$ & $\phi^i_{\mu}$ & $\hat     R(\psi^{i}_\mu)$ & $\eta^i$ \\
    $SU(2)_\RS$-symmetry & $V_\mu{}^{i}_{j}$\\
    $U(1)_{\tilde \RS}$-axial symmetry & $ \tilde A_{\mu}$ \\
\hline
& auxiliary fields & \\
tensor &  $T_{\mu \nu a}$& & \\
spinor &  $\chi^{i}$ &  &  \\
scalar & $ M $ & & \\
  \end{tabular}
\medskip
\end{table}

The gauge fields  ($\omega_{\mu}^{\hat \mu \hat
  \nu}$, $f^{\hat \mu}_{\mu}$, $\phi^{i}_{\mu}$)  associated to the
rotation, the special conformal symmetry and the special conformal
supersymmetry  are
expressed in terms of the other fields from the constraints on
superconformal covariant curvatures $\hat R$ for the fields  $e^{\hat \mu}_\mu, 
\omega^{\hat \mu \hat \nu }_{\mu}, \psi^{i}_\mu$.

It is convenient to use 6d spinorial notations for the spinors of
4d $\mathcal{N}=2$  theories under dimensional reduction. We use index
conventions from appendix  \ref{se:conventions} and chirality
conventions as in \ref{ta:6d-chirality}.
\begin{table}[!h]
  \caption{6d chirality
\label{ta:6d-chirality}
}
  \centering
  \begin{tabular}{ccc}
 & Weyl  multiplet & \\
\hline
 field & variation &  6d chirality\\
   $\psi^i_\mu$ & $\ve^i$ & $+1$  \\
   $\phi^i_\mu$ & $\eta^i$ &  $-1$\\
   $\chi^i$  &   & $+1$ \\
   $T_{\mu \nu a}$ &  & $-1$ \\
\hline
& vector multiplet & \\
  $\lambda^i $ &    & $+1$  \\
  \end{tabular}
\end{table}

To find the action of the  $\mathcal{N}=2$ Poincare supergravity 
interacting with $n_V$ vector multiplets and $n_H$ hypermutiplets one
considers  Weyl multiplet coupled with  $n_V +1$ vector multiplets and
$n_H  + 1$ hypermultiplets and then uses one  vector multiplet and one
hypermultiplet as auxiliary fields  to
gauge fix the non-Poincare superconformal gauge symmetries and to
integrate out non-Poincare supergravity fields. Finally, one gets
the on-shell physical fields of the Poincare $\mathcal{N}=2$
supergravity: the frame $e^{\hat \mu}_{\mu}$,  the gravitino doublet
$\psi_\mu^{i}$ and the graviphoton $A_\mu$ together with $n_v$ vector
multiplets and $n_h$ hypermultiplets.  (See more details in the
diagram \cite{VanProeyen:2000}, page 81).

To construct gauge theories on fixed curved backgrounds
with partially preserved  off-shell supersymmetry the full machinery
described above is not necessary. It is sufficient to consider the
off-shell action and
the supersymmetry transformation  for the vector multiplets and hypermultiplets coupled to the Weyl gravity multiplet and then freeze the fields of the
Weyl multiplet to a supersymmetry invariant background \cite{Festuccia:2011fk,Dumitrescu:2012ha,Hama:2012bg}.

The supersymmetry transformation is a linear superposition  of the
Poincare supersymmetry variation $\ve^i$ and the conformal
supersymmetry variation $\eta^i$. 
 Since variation of bosonic fields in Weyl  multiplet is proportional to
 fermionic fields of Weyl multiplet and they are set to zero in the
 background, the supersymmetric equation is the vanishing variation of
the independent fermions $\psi^i_\mu,  \chi^i$.  The field $\phi^i_\mu$
is  expressed in terms of $\psi^i_\mu$ and  $\chi_i$ through the curvature
 constraints, and the vanishing variation of $\psi^i_\mu$ and
$\chi^i_\mu$ automatically implies vanishing variation of
$\phi^i_\mu$. 

We quote  the variation of gravitino  $\psi^{i}_\mu$ and auxiliary
field $\chi^i$ under the Poincare and conformal 
supersymmetries $\ve^i$ and $\eta^i$ from \cite{supergravity:2012}, page 429. 
The  equations are\footnote{In some $\mathcal{N}=2$ supergravity literature the
 auxiliary scalar field $M$ in Weyl multiplet is denoted $D$. For the
conventions on the Clifford algebra see Appendix
\ref{se:appendix-clifford}; the slash symbol on tensors denotes Clifford
contraction as in equation (\ref{eq:slashed}).}:
\begin{equation}
\label{eq:conformalK}
  \begin{aligned}
  & \delta_{\ve,\eta} {\psi^i_\mu} = 
  D_{\mu} \ve^i - \tfrac 1 {16} \slashed{T} 
\gamma_\mu  \ve^i - \gamma_\mu \eta^i  = 0\\
&  \delta_{\ve, \eta} \chi^i =  - \tfrac {1}{24} [D_{\mu} \slashed{T}] \gamma^\mu \ve^i
+ \tfrac 1 6 (\slashed{F}_V^\RS)^{i}_{j} \ve^j + \tfrac{1}{12} \slashed{T}
\eta^i  + \tfrac 1 2 M  \ve^i  = 0\\
  \end{aligned}
\end{equation}
From the first equation the conformal supersymmetry
parameter $\eta$ can be expressed in
terms of the Poincare supersymmetry parameter $\ve$ as 
\begin{equation}
\label{eq:etaep}
   \eta^i = \tfrac{1}{4} \slashed{D} \ve^i
\end{equation}
where we used (\ref{eq:Tzero}). Later we use 
this relation to substitute  $\eta^i$ with
  $\tfrac {1}{4} \slashed{D} \ve^i$ and vice versa.  The second
  equation, 
 called the \emph{auxiliary}  equation,  can be
transformed using the Lichnerowicz formula for $\slashed{D}^2$
(\ref{eq:Lichner})
and the divergence of the first equation, see appendix  (\ref{eq:1}):
\begin{equation}
\label{eq:conformal2}
  \begin{aligned}
&   D_{\mu} \ve^i - \tfrac 1 {16} \slashed{T} 
\gamma_\mu  \ve^i - \tfrac 1 4 \gamma_\mu \slashed{D} \ve^i   = 0\\
& \slashed{D} \eta  = -\tfrac{1}{2}(
 \tfrac 1 6 R + M) \ve +
  \tfrac{1}{16} [D^{\mu} \slashed{T}] \gamma_\mu \ve  \quad \quad ( =  \tfrac 1 4 \slashed{D}^2 \ve)
  \end{aligned}
\end{equation}
Here  $R$ denotes  the scalar curvature   (\ref{eq:curvatures})  of  the
background metric.

The equations (\ref{eq:conformal2}) are called  \emph{generalized conformal Killing spinor
equations},  and the spinor $\ve$  is called \emph{generalized conformal Killing spinor}.

The generalized conformal Killing spinor equations transform
covariantly with respect to local Weyl transformation
\begin{equation}
  g_{\mu \nu} \mapsto e^{2 \Omega} g_{\mu \nu} 
\end{equation}
 with the weights
\begin{equation}
   \qquad   \ve \mapsto  e^{\frac 1 2 \Omega} \ve, \qquad
M \mapsto e^{-2 \Omega} M,  \qquad  \slashed{T}  \mapsto  e^{- \Omega} \slashed{T}
\end{equation}
Therefore the solutions can be classified by their conformal class. 

The generalized conformal Killing spinor equations,
similarly to the conformal Killing equations, can be rewritten as the
generalized parallel transport equations on the section of 
doubled spinorial bundle
\begin{equation}
\mathcal{D}_{\mu}  \begin{pmatrix}
    \ve^i\\
    \eta^i
  \end{pmatrix} = 0
\end{equation}
for certain $\mathcal{D}_{\mu}$. This representation could be useful to classify the solutions. 

The solution to the generalized conformal Killing equations is
particularly simple for conformally flat background with vanishing
auxiliary  field $T_{\mu \nu a}$, flat $\RS$-symmetry gauge connection
and vanishing auxiliary scalar $M$. In the flat
$\BR^{4}$ coordinates $x^{\mu}$, the solution is simply
\begin{equation}
  \ve^i(x) = \hat \ve^i +  \slashed{x} \hat \eta^i 
\end{equation}
where $\ve^i$ and $\eta^i$ are arbitrary constant spinor parameters
associated with translational and special conformal supersymmetry
respectively. The maximal dimension of the space of solutions to the
parallel
transport equation in a bundle is the rank of this bundle. We see that the conformally
flat background with flat $R$-symmetry connection and vanishing $T_{\mu
  \nu a}$ is maximally supersymmetric.  The $16$
sections are generated by  $8$ components of   $\hat \ve$ and $8$ components
of  $\hat \eta$. 

It would be interesting to find the complete classification of  the solutions
to the generalized conformal Killing equation with various amounts of supersymmetry. 
In this note we will focus on particular backgrounds interesting 
for the localization  of gauge theories.

\subsection{Vector multiplet}
The 4d $\mathcal{N}=2$  vector multiplet  $(A_m, \lambda^i, Y^{ij})$ includes the
gauge field $A_\mu$ and two real scalar fields $\Phi_a$  combined into the reduction of 6d
gauge field $(A_{m}) = (A_{\mu},
\Phi_a)$, the $SU(2)_\RS$-doublet of gaugino fermions
$\lambda^{i}$, and the $SU(2)_\RS$-triplet of auxiliary fields
represented by the matrix $Y^{ij}$ symmetric in $(ij)$. The gaugino
$\lambda^{i}$ is the reduction of the $SU(2)_\RS$-doublet of 6d Weyl spinors of chirality $+1$ for
$\gamma_*^{\mathrm{6d}}$. The spinor fields from the $SU(2)_\RS$-doublet
enter into the Lagrangian and supersymmetry variation holomorphically,
their complex conjugates never appear in the Euclidean formulation of
the theory. 

The supersymmetry variation for the vector multiplet is
\begin{equation}
\label{eq:variation}
  \begin{aligned}
&  \delta A_{m} = \tfrac  1 2 \conj{\lambda^i}  \gamma_m \ve_i \\
&  \delta \lambda^i = - \tfrac 1 4 F_{mn} \gamma^{mn} \ve^i  + 
Y^{i}_{\,\,\, j} \ve^{j} + \Phi_a \gamma^{a} \eta+   \tfrac 1 8 T_{\mu \nu a} \Phi^a \gamma^{\mu  \nu} \ve \\
&  \delta Y^{ij} = - \tfrac 1 2 (\conj{\ve^{(i} } \slashed{D}
\lambda^{j)}) \\
  \end{aligned}
\end{equation}
where there are two extra two terms for $\delta \lambda^i$ 
compared to the standard translational supersymmetry. 
In our conventions the supersymmetry parameters 
 $\ve^i, \eta^i$ are bosonic and  $\delta_{\ve, \eta}$ is fermionic, 
for a field $\phi$ the field $\delta_{\ve, \eta} \phi$ has opposite statistics of $\phi$. 

If $\ve$ and $\eta = \tfrac{1}{4} \slashed{D} \ve$ solve the generalized conformal Killing spinor
equations (\ref{eq:conformal2}),  the supersymmetry transformations (\ref{eq:variation}) closes
off-shell:
\begin{equation}
  \begin{aligned}
& \delta_{\ve, \eta}^2 A_{\mu} = \tfrac 1 4  (\conj \ve \gamma^\nu \ve) F_{\nu \mu} + \tfrac 1 4 [  (\conj \ve \gamma^a \ve)
\Phi_a, D_{\mu}] \\
& \delta_{\ve, \eta}^2 \Phi_{a} = \tfrac 1 4 [ (\conj \ve \gamma^m \ve) D_m, \Phi_a] - \tfrac 1 2 (\conj{\eta}
\gamma_{ab} \ve) \Phi^b + \tfrac 1 2 (\conj{\eta} \ve) \Phi_a \\
&    \delta_{\ve, \eta}^2 \lambda^i = \tfrac 1 4 \left( (\conj \ve \gamma^m \ve) D_m \lambda^i + 
\tfrac{1}{4} D_{\mu} (\conj{\ve} \gamma_\nu \ve) \gamma^{\mu \nu}
\lambda^i \right) -    \tfrac 1 8 (\conj{\eta} \gamma_{ab} \ve) \gamma^{ab}
\lambda^i + \tfrac     {3}{ 4} (\conj{\eta} \ve) \lambda^i + (\conj{\eta}^{(i} \ve^{j)})   \lambda_j \\
& \delta_{\ve, \eta}^2 Y^{ij} = \tfrac 1 4 [ (\conj \ve \gamma^m \ve)  D_{m} Y^{ij}] 
 + (\conj{\eta} \ve) Y^{ij} + (\conj{\eta}^{(k} \ve^{i)}) Y^{j}{}_{k} +(\conj{\eta}^{(k} \ve^{j)}) Y^{i}{}_{k}
  \end{aligned} 
\end{equation}
The variation $\delta_{\ve, \eta}^2$ contains the Lie derivative action
by the 6d reduced vector field
\begin{equation}
  v^{m} =  \tfrac 1 4 (\conj{\ve} \gamma^{m}\ve)
\end{equation}
The scalar components $m \equiv
a$  generate  the gauge transformation by $v^{a} \Phi_a$.
\begin{table}
\caption{The symmetry action of $\delta^2_{\ve, \eta}$}
  \centering
  \begin{tabular}{cc}
  $\delta_{\ve,\eta}^2$ acts by & parameter \\
\hline
$\mathcal{L}_{v}$  & $ v^m =  \tfrac 1 4 (\conj{\ve} \gamma^{m}\ve)$\\
$ SO(2)_{\tilde \RS}$ & $\tilde R^{ab} = -\tfrac 1 2 (\conj{\eta} \gamma^{ab} \ve)$ \\
$SU(2)_{\RS}$  &  $ R^{ij} = (\conj{\eta}^{(i} \ve^{j)})   \lambda_j $ \\
dilatation & $(\conj{\eta} \ve)$
  \end{tabular}
\end{table}

The  Lagrangian of 4d $\mathcal{N}=2$ vector multiplet coupled to the Weyl
gravity multiplet can be found in  \cite{deWit:1983xe,
  deWit:1984px, Mohaupt:2000mj}, or  \cite{supergravity:2012} page 433
 \begin{multline}
\label{eq:action}
   S = -\tfrac{1}{\gym^2} \int \sqrt{g} d^4 x \tr  (\tfrac 1 2 F_{mn} F^{mn}  +  \conj\lambda^i  \gamma^m D_{m}
   \lambda_i + (\tfrac{1}{6}R + M)  \Phi_a \Phi^a -  2 Y_{ij} Y^{ij} +\\
 -F^{\mu \nu} T_{\mu \nu a} \Phi^a 
+ \tfrac{1}{4}  T_{\mu \nu a} T^{\mu \nu b} \Phi^a \Phi_b )
 \end{multline}

Provided $\ve$ and $\eta = \tfrac 1 4 \slashed{D} \ve$ satisfy
(\ref{eq:conformalK}) the action $S$ is invariant under $\delta_{\ve,
  \eta}$ 
\begin{equation}
\label{eq:Sclosed}
  \delta_{\ve, \eta} S =0.
\end{equation}

\section{Generalized conformal Killing spinor}
Presently the complete classification of the solutions to the
generalized conformal Killing spinor equation (\ref{eq:conformalK}) is
not available. We list some known examples. In all these examples the
$U(1)_{\tilde R}$ connection is set to zero,  the
square of the supersymmetry transformation $\delta^2_{\ep}$ generates
isometry transformation and possibly $SU(2)_{\mathrm{R}}$ transformation
but without dilatation and $U(1)_{\tilde{\mathrm{R}}}$ transformation.

\subsection{Topologically twisted theories} 
One simple class of solutions which exists on any smooth 4-manifold  is
the  Donaldson-Witten topological twist \cite{Witten:1988ze}. One
sets $R$-symmetry $SU(2)_{\RS}$ connection to
compensate right component  of the  $Spin(4) = SU(2)_{L}
\times SU(2)_{R}$ spin-connection. In the twisted theory the 8 components of the
4d $\mathcal{N}=2$ spinor generators transform as a  one-form,  self-dual two-form and  scalar. The scalar component
yields the scalar supersymmetry charge defined on any smooth
4-manifold. The theory localizes to the instanton configurations  $F^{+}_A = 0$.

\subsection{Omega background}
 Another example is the equivariant twist of the topologically twisted
 theory on any manifold with $U(1)$ isometry. To construct such theory,
 one uses a combination of the scalar supersymmetry of the topologically
 twisted theory and the  one-form supercharge contracted with the vector
 field that generates $U(1)$ isometry. Localization of such theory on
 $\BR^4$ counts equivariant instantons and gives Nekrasov partition
 function \cite{Nekrasov:2002qd, Nakajima_2003Lectures,
   Moore:1997dj,Losev:1997tp,Lossev:1997bz}, 
with two equivariant parameters $\ep_1, \ep_2$, each associated to the
rotation of the $\BR^{2}$ planes in the decomposition $\BR^{4}_{\ep_1, \ep_2} = \BR^{2}_{\ep_1} \oplus
\BR^{2}_{\ep_2}$. For a review of instanton counting see contribution
\cite{Te} of this volume.

\subsection{Conformal Killing spinor} Another example is conformally flat 
and $SU(2)_R$-flat metric with $T_{\mu \nu a} = 0$ and conformal Killing
spinor. A  spinor of this type has been used to localize the 
physical $\mathcal{N}=2$ gauge theory on  $S^4$
\cite{Pestun:2007rz}. The isometry vector field has two fixed points:
the north and the south poles of $S^4$. In the neighborhood of the north
pole the theory is locally isomorphic to the theory in the
Omega-background with parameters $\ve_1 = \ve_2 = r^{-1}$ where  $r$ is
the radius of $S^4$, counting equivariant instantons $F^{+}_{A} = 0$. In the
neighborhood of the south pole the theory is conjugate to the theory in
Omega-background, and it counts equivariant anti-instantons $F^{-}_{A} =
0$. The complete partition function on $S^4$ is the fusion of the
Nekrasov partition function and its conjugate:
\begin{equation}
\label{eq:ZS4}
  Z_{S^4} = \int [d\mathbf{a}]\,  | Z_{\BR^{4}; r^{-1},r^{-1}}(\ii \mathbf{a})|^2 
\end{equation}
where $Z_{\BR^4; \ep_1, \ep_2}(\mathbf a)$ is the complete partition function in
Omega background including the classical and perturbative factors, and
$\mathbf{a}$ is the gauge Lie algebra equivariant argument of $Z_{\BR^4; \ep_1,
  \ep_2}$ that physically is interpreted as the electrical type
special coordinate on the Coulomb moduli space of the $\mathcal{N}=2$
theory or boundary conditions  at infinity of
$\BR^{4}_{\ep_1; \ep_2}$  for the scalar field $\Phi_a$ of the
vector multiplet. In the formula (\ref{eq:ZS4}) we omit from the
arguments of the partition function the parameters of the Lagrangian. 

More generally, the cases of $\Omega$-background and conformal Killing
spinor could be viewed as the specialization of local $T^2$-bundle
geometry.

\subsection{Local $T^2$-bundles}

Consider a manifold $X_4$ endowed with the metric structure of
 the warped product $X_4 = T^2_{w_1,w_2} \tilde \times  \Sigma_2$ where $\Sigma_2$ is a 
 Riemann surface, possibly with boundaries,  and $T^2_{w_1,w_2}$ is a
 flat 2-torus
 with basis cycles of length $(2 \pi w_1, 2 \pi w_2)$. Here $(w_1, w_2)$ are locally arbitrary
functions on $\Sigma_2$.  This geomery  generalizes the Omega
background on $\BR^{4}$  and the standard conformal Killing spinor
geometry on $S^4$. The  ellipsoid solution \cite{Hama:2012bg} is a
special  example of $X_4$. Another case was studied in \cite{Nosaka:2013cpa}.

We denote the coordinates along the two circles on $T^2$  by $(\phi_1,
\phi_2)$. We pick  two real parameters $(\ep_1 , \ep_2) $ with the aim
to get  $\delta_{\ep, \eta}^2$ action by the vector field
\begin{equation}
  v =  \ep_1  \partial_{\phi_1} +  \ep_2 \partial_{\phi_2}
\end{equation}
We assume that $\ep_i$ are such that  $ ( \ep_1 w_1)^2 + (\ep_2 w_2)^2
\leq 1$ everywhere on $\Sigma$.
For any  generic functions $(w_1, w_2)$ on $\Sigma$ we can always
find  local  coordinates $(\theta, \rho)$ such that
 \begin{equation}
\label{eq:special}
   \begin{aligned}
&   \cot \theta := \frac{ \ep_1  w_1}{ \ep_2 w_2}  \\
&  \sin^2 \rho: = ( \ep_1 w_1)^2 + (\ep_2 w_2)^2  
   \end{aligned} \qquad 
\Leftrightarrow  \qquad 
  \begin{aligned}
  w_1 = \ep_1^{-1} \sin \rho \cos \theta \\ w_2 = \ep_2^{-1} \sin \rho \sin \theta 
  \end{aligned}
\end{equation}
After we have fixed special coordinates $(\theta, \rho)$ on $\Sigma$, the
metric components $g_{\mu \nu}(\rho, \sigma)$ are parametrized by three
arbitrary functions $g_{\theta \theta}(\theta,\rho), g_{\theta \rho}(\theta,\rho)$ and $g_{\rho
  \rho}(\theta,\rho)$.

Next we choose the frame on $X^4$  of the form
\begin{equation}
\label{eq:frame}
  \begin{aligned}
    e^1 &= w_1(\theta, \rho) d \phi_1 \\
    e^2 &= w_2(\theta, \rho) d \phi_2 
  \end{aligned}
\qquad \qquad   \begin{aligned}
    e^3 &=  e^3_\theta (\theta, \rho) d \theta + e^3_\rho (\theta, \rho) d \rho\\
    e^4 &=  e^4_\rho (\theta, \rho) d \rho 
  \end{aligned}
\end{equation}
Three functions $ e^3_\theta(\theta,\rho),  e^3_\rho(\theta,\rho) ,
e^4_\rho(\theta,\rho) $ 
generically parametrize 2d metric by the relations 
 $g_{\theta \theta } = e_{ \theta }^3 e_{\theta}^3$, 
$ g_{\theta \rho} = e_{\theta}^3 e_{\rho}^3$ and $g_{\rho \rho} =
e_{\rho}^3 e_{\rho}^3 + e_{\rho}^4 e_{\rho}^4$. 
It is convenient to denote 
\begin{equation}
  e^3_\theta(\theta, \rho) \equiv  \sin \rho f_1(\theta, \rho) \qquad
 e^3_\rho(\theta, \rho) \equiv  f_3(\theta, \rho) \qquad
 e^4_\rho(\theta, \rho) \equiv  f_2(\theta, \rho)
\end{equation}
and present solution for the background fields $T_{\mu \nu}$ and
$V_{\mu}{}^{i}_{j}$ in terms of $f_1, f_2, f_3$.\footnote{In 
these notations the solution can be easily specialized to the
Hama-Hosomichi ellipsoid   \cite{Hama:2012bg}
metrically defined by the equation in $\BR^5$ with the standard metric
\begin{equation*}
  r_1^{-2} (X_1^2 + X_2^2) + r_2^{-2} (X_3^2 + X_4^2) + r^{-2} X_5^2 = 1
\end{equation*}
 by taking
 \begin{equation*}
   \begin{aligned}
     X_1 + \imath X_2 = r_1 \sin \rho \cos \theta e^{\imath \phi_1}, \quad
     X_3 + \imath X_4 = r_2 \sin \rho \sin \theta  e^{\imath \phi_2}, \quad   X_5 = r \cos \rho 
   \end{aligned}
 \end{equation*}
and 
  \begin{equation*}
    \begin{aligned}
  &  f_1(\theta, \rho) = f_{\mathrm{HH}}(\theta) = \sqrt{ r_1^2 \sin^2
      \theta + r_2^2 \cos^2 \theta} \qquad 
    f_2(\theta, \rho) = g_{\mathrm{HH}}(\theta, \rho) = \sqrt{ r^2
      \sin^2 \rho + 
r_1^2 r_2^2 f_1(\theta)^{-2}  \cos^2 \rho }\\
&    f_3(\theta, \rho) =h_{\mathrm{HH}}(\theta, \rho) = (-r_1^2 + r_2^2)
f_1(\theta)^{-1} \cos\theta
      \sin \theta \cos \rho 
    \end{aligned}
  \end{equation*}
In the case of round sphere $S^4$ we set 
\begin{equation*}
  f_1(\theta, \rho) = r \qquad f_2 (\theta, \rho) = r \qquad  f_3(\theta, \rho)
  = 0
\end{equation*}
 }

In the   $\gamma$ matrix basis (\ref{eq:standard4d}) we choose the
$SU(2)_\RS$-doublet spinor $(\ve^1, \ve^2)$ in the frame  (\ref{eq:frame}) to be given by
 \begin{equation}
\label{eq:spinor}
   \begin{aligned}
&   \ve^{1} = e^{\frac 1 2( i \phi_1 + i \phi_2)} 
( e^{- i \frac \theta 2} \sin \tfrac \rho 2 , 
 - e^{ i \frac \theta 2} \sin \tfrac \rho 2, 
i e^{-i \frac \theta 2} \cos \tfrac \rho 2,
-i e^{i \frac \theta 2 } \cos \tfrac \rho 2) \\
& \ve^{2} = 
e^{- \frac 1 2( i \phi_1 + i \phi_2)}
( e^{ - i \frac \theta 2} \sin \tfrac \rho 2,
 e^{ i \frac \theta 2} \sin \tfrac \rho 2,
-i e^{-i \frac \theta 2 } \cos \tfrac \rho 2,
-i e^{i \frac \theta 2 } \cos \tfrac \rho 2)
   \end{aligned}
 \end{equation}
Notice that this spinor satisfies the standard reality condition
$\ve^2 = c  \bar \ve^1$ where bar is the complex conjugation and $c$ is the Majorana bilinear matrix
(\ref{eq:standard4d}), and that this spinor is  the
transformation to the $(\phi_1, \phi_2, \rho, \theta)$ frame of the 
standard conformal Killing spinor $\ve(x) = \ve_s + x^\mu \gamma_\mu
\ve_c$ that was used in \cite{Pestun:2007rz} for $S^4$, where $x^\mu$
are stereographic projection coordinates on $S^4$. 

For the spinor (\ref{eq:spinor}) we find the bilinear vector field\footnote{In the equation (\ref{eq:bilvector}) $\ve^i$ denotes the
  $+1$ chiral $6d$ spinors and $\gamma^m$ for the $6d$ gamma-matrices, 
while in the equation (\ref{eq:spinor}) the components of
the spinor $\ve^{i}$ are presented  
with respect to the 4d Clifford algebra representation (\ref{eq:standard4d}).}
\begin{equation}
\label{eq:bilvector}
 v^m = \tfrac 1 4 \ve^i \gamma^m \ve_i = \tfrac 1 2 \ve^1 \gamma^m
 \ve^2 :     \quad \quad 
  \begin{aligned}
&  v^\mu|_{\mu \in (\phi_1,  \phi_2, \theta, \rho)} = (\ep_1, \ep_2, 0, 0)\\
& v^a|_{a \in (5,6)}  = (- \cos \rho, -i )
  \end{aligned}
\end{equation}
This vector field is the natural isometry of the $T^2$-bundle $X_4$.

Under  the ansatz (\ref{eq:spinor}), the equations on the background
fields $V_\mu{}^i_j, T_{\mu \nu}, M$ are inhomogeneous ordinary linear  equations,
which  can be directly solved. Though the system is overdetermined,
as there are $32+8$ linear equations from $\delta \psi^i_\mu$ and  from $\delta
\chi^i$  on $12 + 6 + 1=19$ components for $V, T, M$, we find that solution
always exists for any $T^2$-bundle. Moreover, the solution is not
unique; the space of solutions forms a vector bundle of rank  three.  This is completely analogous to the case of the
Hama-Hosomichi ellipsoid \cite{Hama:2012bg}. 

Below $T_{\hat \mu \hat \nu}$ denote the components of $T$ in the frame (\ref{eq:frame})
$e^{\hat \mu}_{\mu}$,   so that $T_{ \mu  \nu} = T_{\hat \mu \hat \nu} e^{\hat \mu }_{\mu}
  e^{\hat \nu }_{\nu}$. The $V$ denotes the connection one-form  of the $SU(2)_{\RS}$ gauge field 
$D = d + V$.  The components of the $T$ do not depend on
$(\phi_1, \phi_2)$, and the components of  $V =  i \sigma_I
V^{I}$,  where  
$\sigma_I$ are the standard Pauli matrices,   depend on $(\phi_1, \phi_2)$
as
\begin{equation}
\label{eq:V-field}
  \begin{pmatrix}
  V^1 \\
  V^2
  \end{pmatrix} =
  \begin{pmatrix}
    \cos(\phi_1 + \phi_2) & \sin(\phi_1 + \phi_2) \\
    -\sin(\phi_1 + \phi_2) & \cos (\phi_1 + \phi_2)
  \end{pmatrix}
  \begin{pmatrix}
    \hat V^1\\
 \hat V^2
  \end{pmatrix}, \qquad V^3 = \hat V^3 
\end{equation}
where $\hat V$ is constant in  $(\phi_1, \phi_2)$.

The particular solution is 
\begin{equation}
\label{eq:Tsolution}
  \begin{aligned}
&    T_{12} = 0  & \qquad &T_{34} = 0 \\
  &  T_{13} = 2 \sin \theta \left ( \frac{1}{f_1} - \frac{1}{f_2} \right)
&\qquad      &  T_{23} =-2 \cos \theta \left ( \frac{1}{f_1} -
      \frac{1}{f_2} \right) \\
&T_{14} = - \frac{ 2 \sin \theta f_3}{ f_1 f_2} &  \qquad 
&T_{24} = \frac{ 2 \cos \theta f_3}{f_1 f_2} 
\end{aligned}
\end{equation}
for $T$ components and 
\begin{multline}
\label{eq:Vsolution}
\hat V  =  \left(- \frac {1}{4 \ep_1} \sin 2 \theta \cos \rho \left( \frac{1}{f_1} -
  \frac{1}{f_2}\right) + \frac{ \sin^2 \theta f_3 }{2 \ep_1 f_1 f_2}
\right) i \sigma_2  \, d \phi_1 + \\
\qquad \left( \frac{1}{4 \ep_2} \sin 2 \theta \cos \rho \left (\frac 1 {f_1} -
  \frac 1 {f_2} \right) + \frac{ \cos^2 \theta f_3}{ 2 \ep_2 f_1 f_2}\right) i
\sigma_2 \, d \phi_2  + \\
\qquad \left(  -\frac 1 2 + \frac { \sin^2 \theta }{ 2 \ep_1 f_1} + 
\frac {\cos^2 \theta r_1}{ 2 \ep_1 f_2 } + \frac{ \sin 2 \theta \cos \rho f_3
}{4 \ep_1 f_1 f_2} \right) i \sigma_3 \, d \phi_1 + \\
\qquad \left( -\frac 1 2 + \frac{ \cos^2 \theta }{ 2 \ep_2 f_1}
+ \frac{ \sin^2 \theta }{2 \ep_2 f_2} - \frac { \sin 2 \theta \cos \rho f_3
}{ 4 \ep_2 f_1 f_2} \right) i \sigma_3 \, d \phi_2 + \\
\qquad \left (  \frac{  (f_1 - f_2) \cos \rho + \sin
  \rho \partial_\rho f_1 - \partial_\theta f_3  }{ 2 f_2}  \right) i
\sigma_1 \, d \theta\\
+ \left( \frac{f_3 \left( \sin\rho \, \partial_\rho f_1 +  f_1 \cos \rho -
   \partial_{\theta} f_3  \right)  -   f_2 \partial_{\theta} f_2 }
{2    f_1 f_2 \sin \rho }  \right) i \sigma_1 \, d \rho 
\end{multline}
are the $V$ components. 
This particular solution can be deformed by three-parametric family 
\begin{equation}
\label{eq:T-deformed}
  \begin{aligned}
&    \delta T_{12} = c_3  &    &\delta T_{34} = c_3 \cos \rho  \\
&  \delta T_{13} = - c_1 \cos \rho \sin \theta  - c_2 \cos\theta &
&\delta T_{23} =  c_1 \cos \rho \cos\theta - c_2 \sin \theta \\
& \delta T_{14} = c_1 \cos \theta - c_2 \cos \rho \sin \theta & & \delta
T_{24} = c_1 \sin \theta + c_2 \cos \theta \cos \rho \\
\end{aligned}
\end{equation}
together with
\begin{multline}
\label{eq:V-deformed}
  \delta \hat V = \left(  -\frac 1 4 c_1  \sin^2 \rho f_1 d\theta -\frac 1 4
\sin\rho c_1  f_3 d \rho - \frac 1 4 c_2 \sin \rho f_2 d \rho \right) i
\sigma_1 + \\
\left (  -\frac 1 {8 \ep_1} c_1 \sin 2 \theta \sin^2\rho \,
d\phi_1 
+ \frac 1  {8 \ep_2} c_2 \sin 2 \theta \sin^2 \rho  d\phi_2 - \frac 1 4
c_3 \sin \rho f_2  d \rho \right ) i \sigma_2 + \\
\left(- \frac 1 {8 \ep_1} c_2  \sin 2 \theta \sin^2 \rho
  d\phi_1 + \frac 1  {8 \ep_2}  c_2  \sin 2 \theta \sin^2 \rho d\phi_2+ \frac 1
4 c_3  \sin^2 \rho f_1 d \theta + \frac 1 4 c_3 \sin \rho f_3 d\rho
\right) i \sigma_3 
\end{multline}
where $c_1, c_2, c_3$ are arbitrary functions on $\Sigma$. 
The background auxiliary scalar $M$ is 
\begin{multline}
\label{eq:Mscalar}
  -\frac 1 2 ( \frac 1 6 R + M) = \frac{1}{ 4 f_1^2} - \frac{1}{4 f_2^2} - \frac{1}{ f_1 f_2} + \frac{
    f_3^3}{4 f_1^2 f_2^2}   \\
+ c_1 \left ( 
-  \frac { \cos \rho }{4 f_1} + \frac {3  \cos \rho}{ 4 f_2}  - \frac{
  \cot 2 \theta f_3}{ 2 f_1 f_2} + \frac { \sin \rho \partial_\rho f_1 }
{ 4 f_1 f_2} - \frac{ \partial_\theta f_3} {4 f_1 f_2} \right)  + \frac{ \sin \rho} {4 f_2} \partial_\rho c_1 
- \frac {  f_{3}}{ 4 f_1 f_2} \partial_\theta c_1   \\
+ c_2 \left (
 - \frac{ \cot 2 \theta}{ 2 f_1}
+ \frac { \cos \rho f_3}{4 f_1 f_2}
- \frac { \partial_\theta f_2 }{ 4 f_1 f_2} \right) 
- \frac{1}{4 f_1} \partial_\theta c_2  - \frac{1}{16} \sin^2 \rho \left ( c_1^2 + c_2^2 + c_3^2 \right) 
\end{multline}

\subsection{Four-sphere} A topological four-sphere $X_4 = S^4_{\ep_1,\ep_2}$
with $T^2$ invariant metric can be presented as a local $T^2_{w_1,w_2}$ 
bundle fibered over a two-dimensional digon $\Sigma_2$. One of the
cycles of $T^2_{w_1,w_2}$ collapses at one edge of the digon, and the other
cycle  collapses at the other edge:
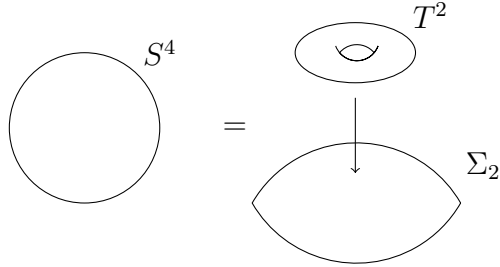
\begin{center}
\begin{tikzpicture}
  \draw (0,0) circle [radius=1cm] [xshift=1cm,yshift=1cm] node {$S^4$}   [xshift=1cm,yshift=-1cm] node {$=$};
  \draw (5,-1) arc (30:150:16mm) arc (210:330:16mm) [xshift=3mm,yshift=5mm] node {$\Sigma_2$};
  \draw (3.6,1)  ellipse (8 mm and 4 mm)  [xshift=10mm,yshift=5mm] node {$T^2$};
  \draw (3.85,1) arc (40:140:3mm)  arc (220:340:3mm) arc (340:200:3mm);
  \draw [->] (3.6, 0.4) -- (3.6,-0.6);
\end{tikzpicture}
\end{center}

The  coordinates $(\theta,
\rho)$ on the base $\Sigma_2$ are in  the range $(\theta,\rho) \in [0,\frac {\pi}{2}] \times [0,
\pi]$. The $w_1$ cycle collapses at $\theta = \frac{\pi}{2}$ and the
$w_2$ cycle  collapses at $\theta = 0$. Both circles collapse in the corners of the digon. The corner $\rho = 0$ will be called
the north pole, and the corner $ \rho = \pi$ will be called the south
pole. The metric on $S^4_{\ep_1,\ep_2}$ is smooth at the cusps and the
edges of the digon $\Sigma_2$  if the functions $f_i(\theta, \rho)$
satisfy asymptotically 
\begin{equation}
  \begin{aligned}
&f_1(\theta,\rho)|_{\theta =0} = \ep_2^{-1}, \qquad  \qquad f_1(\theta,\rho)|_{\theta = \frac \pi 2} =  \ep_1^{-1} \\
& f_1(\theta,\rho)|_{\rho = 0, \pi} =  ( \ep_1^{-2} \sin^2 \theta +
   \ep_2^{-2} \cos^2 \theta)^{\tfrac 1 2} \\
& f_3(\theta, \rho)|_{\rho = 0, \pi}  = \pm (\ep_2^{-2} - \ep_1^{-2})f_1(\theta,\rho)^{-1} \cos \theta \sin \theta  \\
& f_2(\theta, \rho)|_{\rho = 0, \pi} =  (\ep_1 \ep_2 f_1(\theta,\rho))^{-1}
  \end{aligned}
\end{equation}

The metric along $\Sigma_2$ is arbitrary in the interior. In
particular, taking $f_2(\theta, \rho)$ very large in the interior, it is possible
to stretch $S^4_{\ep_1,\ep_2}$ to a very long cylinder with two hemispherical caps
attached at the ends. Localization on this
geometry presumably is related to the convolution of the ground state topological wave
functions with its conjugate by cutting the $S^4_{\ep_1,\ep_2}$ in the middle at $\rho
=\frac \pi 2$, as in the AGT correspondence \cite{Alday:2009aq} with Liouville
theory and quantum Teichmuller theory \cite{Te,Vartanov:2013ima}, and Nekrasov-Witten construction
\cite{Nekrasov:2010ka}. 

The background fields (\ref{eq:Tsolution})(\ref{eq:Vsolution})
and the spinor (\ref{eq:spinor}) with generic smooth
functional parameters  $c_1, c_2, c_3$
(\ref{eq:T-deformed})(\ref{eq:V-deformed})
is a supersymmetric background only in the
interior of $\Sigma_2$. 
The coordinates are singular at the north and the south poles  $\rho =
0$ and $\rho = \pi$. We need to ensure that the spinor $\ep$ and 
 the background fields $V$ and $T$ are smooth in a proper coordinate
 system around the poles. At the north pole $\rho =0$ we choose approximately Cartesian
coordinates
\begin{equation}
\begin{aligned}
  x_1 = 2 \ep_1^{-1}  \tan {\tfrac \rho 2} \cos \theta \cos \phi_1\\
  x_3 =2 \ep_2^{-1}  \tan {\tfrac \rho 2} \sin \theta \cos \phi_2 
\end{aligned} \qquad \qquad 
\begin{aligned}
  x_2 = 2 \ep_1^{-1}  \tan {\tfrac \rho 2}  \cos \theta \sin \phi_1 \\
  x_4 = 2 \ep_2^{-1}  \tan {\tfrac \rho 2}  \sin \theta \sin \phi_2 
\end{aligned}
\end{equation}
with the standard frame $e^{\hat \mu}_\mu  = \delta^{\hat
  \mu}_\mu$.  In the $x$-frame the spinor (\ref{eq:spinor}) becomes
\begin{equation}
 ^{x}\ve = e^{ - \frac \pi 4 \gamma_{12}} e^{ - \frac{\phi_1}{2}
     \gamma_{12} - \frac{\phi_2}{2} \gamma_{34}} e^{ - \frac{\pi}{4}
     \gamma_{24}} e^{ \frac{\beta}{2} \gamma_{34}} \ve , 
   \qquad \text{with} \qquad \scriptstyle {\sin \beta = \frac{  \ep_1^{-1} \sin \theta}{ \ep_1^{-2} \sin^2
  \theta + \ep_2^{-2} \cos^2 \theta}}
\end{equation}
The spinor ${^x}\ve$ is not smooth in the $x$-frame but is $SU(2)_\RS$
gauge equivalent to the conformal class of the
standard smooth spinor in the Omega-background \cite{Nekrasov:2002qd} (up to the Weyl transformation $
^{x} \ve_{\Omega} \to {^{x} \ve_{\Omega}} \cos \tfrac \rho 2$):
\begin{equation}
  \begin{aligned}
  ^{x} \ve_{\Omega} :=  (\hat \ve_{s}  - \tfrac{1}{2} \Omega_{\mu \nu} x^{\mu}
  \gamma^{\nu} \hat \ve_s ) \cos \tfrac{\rho}{2} 
  \end{aligned}
\end{equation}
where non-zero components of $\Omega$ are $\Omega_{12} = -\Omega_{12} = \ep_1$ and
$\Omega_{34} = - \Omega_{43} = \ep_{2}$. Namely, for $\hat \ve_s =
(1+i)(0,0,1,0)$ we
find 
\begin{equation*}
  \begin{aligned}
& ^{x} \ve_{\Omega}^{1} = (1+i)(\cos \tfrac \rho 2) ( -  \tan \tfrac \rho 2 \sin
 \theta e^{i \phi_2}, 
- \tan \tfrac \rho 2  \cos \theta e^{i \phi_1}, 1 ,  0) \\
& ^{x}\ve^{1} = (1+i)(\cos \tfrac \rho 2)  ( -  \tan \tfrac \rho 2 \sin
\tfrac { \beta +  \theta}{2}  e^{i \phi_2}, 
- \tan \tfrac \rho 2  \cos  \tfrac{ \beta + \theta}{2} e^{i \phi_1},  \cos \tfrac{ \beta - \theta}{2}  ,  -\sin \tfrac { \beta - \theta}{2}  e^{i
  (\phi_1 + \phi_2)})
  \end{aligned}
\end{equation*}
with $\ve^{2}$ found by Majorana conjugation $\ve^2 = c \bar
\ve^1$. 

The $SU(2)_\RS$ gauge transformation relating spinor
$^{x} \ve$  (\ref{eq:spinor}) and the Omega-background spinor
$\ve_{\Omega}$  near the north pole is 
\begin{equation}
 \ve^{i}_{\Omega} = U^{i}_{j} \ve^{j}, \qquad  \qquad U = 
\left( \begin{smallmatrix}
    \cos \tfrac { \theta - \beta}{2} & i e^{i (\phi_1 + \phi_2)} \sin
    \tfrac {\beta - \theta}{2} \\
    i e^{-i(\phi_1 + \phi_2)} \sin \tfrac {\beta -\theta}{2} & 
\cos \tfrac { \theta - \beta}{2} 
 \end{smallmatrix}\right)
\end{equation}

Requiring that $SU(2)_\RS$ gauge field $^{U} V = U d U^{-1} + U V U^{-1}$ (\ref{eq:V-field})(\ref{eq:Vsolution})(\ref{eq:V-deformed}) is smooth at the origin, 
and that components  $T_{\mu \nu}$ are well defined in the $x$-frame, we
find the parameters
\begin{equation}
\label{eq:parameters}
  \begin{aligned}
     c_1 =\left( \frac{1}{f_1} - \frac{1}{f_2} \right) \varphi(\rho), \quad 
     c_2 = - \frac{ f_3}{  f_1 f_2} \varphi(\rho) , \quad 
     c_3 = 0
  \end{aligned}
\end{equation}
where $\varphi(\rho)$ is any smooth function such that  $
\varphi(\rho)_{\rho=0} = 1 + O(\rho^2)$ and $\varphi(\rho)_{\rho =  \pi} =
 -1 + O(\rho^2)$. Then the gauge field $^U V$ is smooth everywhere and
 $T_{\rho=0} = T^{-}$, $T_{\rho = \pi}  = T^{+}$. 

In our conventions the spinor $\ve$ is of positive chirality  at the
north pole (transforms under self-dual spacial rotations) and
negative-chirality at the south pole (transforms under anti-self-dual rotation). In the zeroth
order approximation the theory around the north pole is topological
Donaldson-Witten theory that localizes to configurations $F^{+}=0$, and 
the theory around the south pole is conjugated and localizes to
configurations $F^{-} =0$.  In the first order approximation the theory
around poles is equivalent to the theory in the Omega-background, and
localizes respectively to the equivariant instantons $F^{+} = 0$ at the
north pole and equivariant anti-instantons $F^{-} = 0$ around the south
pole. 

With the choice (\ref{eq:parameters}) at $\rho = 0$ we find that
non-zero components of $T = T^{-}$ are
\begin{equation}
  T_{12} = -T_{34} = \ep_{1} - \ep_2
\end{equation}
If the geometry in the neighborhood of the north pole is approximated by
the embedded ellipsoid in $\BR^{5}$  with radia $(r_1,r_1,r_2,r_2,r)$
for  $r_1=\ep_1^{-1}, r_2=\ep_2^{-1}$ as in \cite{Hama:2012bg} then 
the curvature of the $SU(2)_{\RS}$ background field at $\rho=0$ is
particularly simple 
\begin{equation}
\begin{aligned}
F_V & = \left(\frac{-2 r^2+r_1^2+r_2^2}{4 r_1^2 r_2^2} (dx_1 \wedge dx_3 -
  dx_2 \wedge dx_4) \right)  i
\sigma_1   \\
&+\left(-\frac{-2 r^2+r_1^2+r_2^2}{4 r_1^2 r_2^2} (dx_1 \wedge
  dx_4 + dx_2 \wedge dx_3) \right) i \sigma_2  \\ 
 &+ \left( \frac{r_1^2-r^2}{2 r_1^4} dx_1 \wedge dx_2 +
\frac{r_2^2-r^2}{2 r_2^4} dx_3 \wedge dx_4 \right)  i \sigma_3
\end{aligned}
\end{equation}
One can compare $F_V$ with the metric curvature at the north pole and
notice the difference: the $SU(2)_{\RS}$ background field differs from the
usual topologically twisted theory. The non-zero metric curvature
components in the $x$-frame at the north pole are
\begin{equation}
  R^{12}{}_{12} = \frac {r^2} {r_1^2}, \qquad 
  R^{34}{}_{34} = \frac{r^2}{ r_2^2}, \qquad
  R^{24}{}_{24} = R^{13}{}_{13} = R^{14}{}_{14} = R^{23}{}_{23} =\frac{r^2}{r_1^2 r_2^2} 
\end{equation}

\subsection{Superconformal Index} 
For this geometry the base is the product of an interval and the circle  $\Sigma_2 = I_{\la \theta \ra} \times S^{1}_{\la \rho
   \ra}$. At the ends
 of the interval $I$ the two circles of $T$ collapse. The slice of $X^4_{\ep_1,\ep_2}$
 at fixed $\rho$ is topologically an $S^3_{\ep_1,\ep_2}$, and then $X_4 = S^3_{\ep_1,\ep_2} \times
 S^1$. A suitable $SU(2)_\RS$ background field ensures existence of
 unbroken supercharge.  The partition function on $S^3_{\ep_1,\ep_2} \times
 S^1$ computes the superconformal
 index
 \cite{Kinney:2005ej,Romelsberger:2005eg,Dolan:2008qi,Dolan:2011rp,Gadde:2011uv}
 and \cite{H} section 4.1 and \cite{RR} in this volume. 

\subsection{Other geometries}
It would be interesting to study more general four-manifolds with the
structure of local $T^2$ bundle such as  $S^2 \times
S^2$ or $T^2 \times \Sigma_2$ where $\Sigma_2$ is a Riemann surface.

\section{Localization}

Often a supersymmetric quantum field theory with a particular choice of
the supercharge can be interpreted as infinite-dimensional version of
the Cartan model for $\mathcal{G}$-equivariant cohomology on the space
of fields of the theory, see
e.g. \cite{Witten:1988ze,Witten:1992xu}. The supercharge $Q$ plays the role of
the equivariant differential. The path integral is
interpreted as the infinite-dimensional version of
Mathai-Quillen form for the Thom class of the BPS equations bundle over
the space of fields \cite{Atiyah:1990,Cordes:1994fc}. For example, 
 in the Donaldson-Witten topological gauge theory \cite{Witten:1988ze}, 
the space of fields is the infinite-dimensional affine space of
connections $\mathcal{A}$ in a given
principal $G$-bundle on a four-manifold $X_4$ for a compact Lie group $G$, the group $\mathcal{G}$ of the equivariant
action is the infinite-dimensional group of gauge transformations, and
the fibers of the equation bundle over $\mathcal{A}$ is the
space of self-dual adjoint valued two-forms. The Mathai-Quillen form for
the Thom class, with a choice of section $F^{+}_A$, localizes to the
zeroes of the section: instanton configurations. 
The  construction is equivariant with respect to the $\mathcal{G}$
action on $\mathcal{A}$. The path integral over
$\mathcal{A}/\mathcal{G}$ 
reduces to the integration over the
instanton moduli space $\mathcal{M_{\mathrm{inst}}} = \{ A | F_{A}^{+} =0\}/\mathcal{G}$.

\subsection{Omega background} See  \cite{Nekrasov:2002qd, Nakajima_2003Lectures,
   Moore:1997dj,Losev:1997tp,Lossev:1997bz} and \cite{Ta} in this volume.   A conventional 4d $\mathcal{N}=2$ theory with Lagrangian formulation
  is specified by the choice of a compact semi-simple  Lie group $G$ for the gauge
  group and a representation $R$ of $G$ for the hypermultiplet
  matter. The automorphism group of the representation $R$ is the flavor group $\Gf$.
The path integral of $\mathcal{N}=2$ theory in Omega background
$\BR^{4}_{\ep_1, \ep_2}$  localizes to
the equivariant form on moduli space of instantons, further integration
over moduli space is localized to the  fixed points the equivariant group
 action. The equivariant group $\Gequiv = \GL \times G \times \Gf$ is the product of the
isometry of the space-time $\GL = SO(4)$, the gauge group $G$ that acts
 on the  framing at infinity, and the flavour group
$\Gf$. Let  $\Tequiv$ be the maximal torus of the equivariant group 
$\Tequiv = T_{\GL} \times T_G \times T_{\Gf}$. 
The coordinates on the complexified Lie 
algebra of $\Tequiv$ are $(\boldsymbol{\ep}, \mathbf{a}, \mathbf{m} )$. Physically, the parameters 
$\mathbf{a}$  are the asymptotics at the space-time infinity 
of the scalar field $\Phi$ in the gauge vector multiplet, the parameters $\mathbf{m}$
are the matter fields masses, and the parameters $\boldsymbol{\ep} = (\ep_1,\ep_2)$ are the
equivariant space-time rotation angular momenta, the $\Omega$-background parameteres. In our 
conventions the subscript $\Tequiv$ denotes the dependence on
$(\boldsymbol{\ep}, \mathbf{a}, \mathbf{m} )$.

The  partition function $Z$ in the Omega background can be represented as a product of the
classical, perturbative and non-perturbative contributions:
\begin{equation}
\label{eq:total}
 Z_{\Tequiv}(\mathbf{q})   =   Z^{\rm  tree}_{\Tequiv}  Z^{\rm 1-loop}_{\Tequiv}
 Z^{\rm inst}_{\Tequiv} (\mathbf{q})
 \end{equation}  

Formally, 
\begin{equation}
  Z_{\Tequiv}({\bf q})  = \sum_{\mathbf{k}} \mathbf{q}^{k}
  \int_{\mathcal{A}/\mathcal{G}_{\text{gauge}}} \eu_{{\Tequiv}}(\Omega^{2+}
  \otimes \gg) \, \,
  \eu_{{\Tequiv}} ( (S^{-} \ominus S^{+}) \otimes R)
\end{equation}
where $\CalA$ is the infinite-dimensional space of $\Gg$-connections on a principal $\Gg$-bundle
$E \to M$ with fixed trivialization at infinity, 
 $\mathcal{G}_{\text{gauge}} = \Aut(E)$ is the group of gauge
transformations equal to identity at the space-time infinity,
$\Omega^{2+} \otimes \gg$ is the infinite-dimensional vector
bundle over
$\mathcal{A}$ with the fiber being the space of self-dual
$\gg$-valued two-forms, $S^{\pm} \otimes R$ is the
infinite-dimensional vector bundle over $\mathcal{A}$ with the fiber being the space of positive/negative
chirality $R$-valued spinors.

Mathematically, the instanton partition function is
\begin{equation}
\label{eq:Z-inst}
  Z_{{\Tequiv}}^{\rm inst}(\mathbf{q}) = \sum_{{\bf{k}}=0}^{\infty}
  {\bf{q}}^{\bf{k}} \int_{\CalM_{\bf{k}}} \eu_{\Tequiv} (\CalE_{R}).
\end{equation}
Here we are assuming that $\Gg=\times_{i \in I} G_{i}$ where $G_{i}$
are simple factors and $I$ denotes the set of labels for the simple
gauge group 
factors, $\mathbf {q} = \{ q_i| i \in I\}$ is the $|I|$-tuple of the 
exponentiated complexified gauge coupling constants $q_i = \exp (2 \pi \ii \tau_i)$, 
$\mathbf k= \{k_i| i \in I \}$ is an $n$-tuple of non-negative
integers, $k_i$ is the instanton charge (second Chern class)\footnote{For a generic compact simple Lie group $G$
   the integer $k$ classifies the topology of $G$-bundle on $S^4$ by 
 $\pi_{3}(G) = \BZ$. The instanton number $k$ can be computed as 
 \begin{equation*}
   k =  \frac{1}{8 \pi^2} \int_{M}  \langle F, \wedge F\rangle = - \frac{1}{16
     \pi^2 h^{\vee}}  \int_{M} 
\Tr_{\mathrm{adj}} F \wedge F
 \end{equation*}
in the conventions where $F$ is $\gg$-valued two-form, $\langle ,
\rangle$ is the
invariant positive definite bilinear form on $\g$ induced from the
standard bilinear form on $\h^{*}$ in which long roots have length
squared $2$, the $\Tr_{\mathrm{adj}}$ is the trace in adjoint
representation, and $h^{\vee}$ is the dual Coxeter number for $\gg$. For
$G = SU(n)$ the instanton charge $k$  is the second Chern class $k = c_2$. 
} of $G_i$-bundle on the
space-time $M= \BR^{4} = {\mathbf S}^{4} \backslash \infty = {\BC\BP}^{2}\backslash {\BC\BP}^{1}_{\infty}$ with fixed framing at infinity, 
the $\CalM_{\mathbf{k}} = \times_{i  \in I} \CalM_{G_i, k_i}$ is the
instanton moduli space:  $\CalM_{G_i,k_i}$ is 
moduli space of the anti-self-dual $G_i$-connections on $\BR^4$ with the
 second Chern class $k_i$.
The integration measure $\eu_{\Tequiv}(\CalE_R)$ is the ${\Tequiv}$-equivariant Euler class of the matter
bundle $\CalE_{R} \to \CalM_{\bf k}$ where a 
fiber of $\CalE_{R}$ is the space 
of the virtual zero modes for the Dirac operator: $\Gamma(S^{-}\otimes R) \to
\Gamma(S^{+} \otimes R)$ associated to the hypermultiplet. 

Here the classical contribution is 
\begin{equation}
\label{eq:classical}
 Z^{\rm tree}_{{\Tequiv}}(\mathbf{q})= {\exp}\left( - \frac{1}{2 {\ep}_{1}{\ep}_{2}}  
\sum_{i \in \Ver}  2 \pi \ii \tau_i   \la a_i, a_i \ra \right )
\end{equation}
where $\la \ra$ is the standard bilinear form on the Lie algebra of
$G_i$ normalized such that the long root length squared is 2, and
$\tau$ is the complexified coupling constant
\begin{equation}
\label{eq:tau}
  \tau = \frac{ 4 \pi \ii}{\gym^2} + \frac{ \theta}{2 \pi}
\end{equation}

Let
\begin{equation}
R = \bigoplus_{\ell} R_{\ell} \otimes M_{\ell}
\end{equation}
be the decomposition of the matter representation onto the
irreducible, with respect to $G$, components, with the multiplicity
spaces $M_{\ell}\simeq \BC^{N^{\mathrm{f}}_\ell}$, 
on which the masses have the value $m_{\ell,1}, \dots, m_{\ell, N^{\mathrm{f}}_\ell}$. 

The one-loop contribution is expressed in terms of the special function
related to Barnes double gamma function 
\begin{equation}
  G_{\ep_1,\ep_2}(x) = \mathrm{Reg}[\prod_{n_1, n_2 \geq 0}(x + n_1 \ep_1 + n_2 \ep_2)]
\end{equation}
where $\mathrm{Reg}[]$ denotes regularization of the infinite product
with Weierstrass multipliers. 

We find the one-loop factors for the theory
in the Omega background for vector multiplet and hypermultiplet to be
given by
\begin{equation}
\label{eq:1-loop}
  \begin{aligned}
&  Z^{\mathrm{1-loop; vec}}_{\Tequiv}(a;m) = \prod_{i} \prod_{ \alpha \in \Delta^{+}_i}
  G_{\ep_1, \ep_2}(\alpha\cdot a_i) G_{\ep_1,\ep_2}(\ep_1 + \ep_2 - \alpha
  \cdot a_i)\\
&  Z^{\mathrm{1-loop; hyper}}_{\Tequiv}(a;m) = \prod_{\ell} \prod_{\mathfrak{f}=1}^{N_\ell^{\mathrm{f}}}\prod_{ w \in P(R_l)} G_{\ep_1,\ep_2}(
w \cdot a_i + m_{l \mathfrak{f}} + \tfrac 1 2 (\ep_1 + \ep_2))^{-1}
  \end{aligned}
\end{equation}
Here $\Delta^{+}_i$ denotes the set of positive roots for the $i$-th
gauge group factor $G_i$ and $P(R_\ell)$ denotes the set of weights for the
irreducible representation $R_{\ell}$.
These expessions follow from the equivariant index computation by Atiyah-Singer
formula for the self-dual complex and the Dirac complex
respectively. The Aityah-Singer formula for the equivariant index of
complex $C$ evaluated at the group element $g \in \Gequiv$ 
\begin{equation}
  \ind(C;g) = \sum_{f \in \text{fixed points}} \frac { \tr_{C_f}(g)}{ \det_{T_f}(1 - g)}
\end{equation}

 For the self-dual complex 
\begin{equation}
  \Omega^{0} \stackrel{d}{\to} \Omega^{1} \stackrel{d}{\to} \Omega^{2+}
\end{equation}
on $\BR^{4} \simeq \BC^2_{\langle z_1, z_2 \rangle}$ under the
equivariant action $ z_1 \to t_1 z_1, z_2 \to t_2 z_2$ we
find that each two-dimensional weight space of root $\alpha$ and its
conjugate $-\alpha$ contributes as
\begin{equation}
  \frac { \omega+\bar \omega+ t_1 t_2 \omega + \bar t_1 \bar t_2 \bar \omega - ( t_1 \omega +
    \bar t_1 \bar \omega + t_2 \omega + \bar
    t_2 \bar \omega)}{
(1 - t_1)(1 - \bar t_1)(1 - t_2)(1 - \bar t_2)} = \bar \omega \frac{1}{(1
- \bar t_1)(1 - \bar t_2)}  + \omega \frac {1}{ (1 - t_1) (1 - t_2)}
\end{equation}
where 
\begin{equation}
  \omega = e^{ \ii \alpha \cdot a}\quad t_1 = e^{i \ep_1} \quad t_2 =
  e^{i \ep_2}
\end{equation}
Taking $\bar t_1 = t_1^{-1}, \bar t_2 = t_2^{-1}$ and expanding the index
in positive powers of $(t_1,t_2)$ one finds the Chern character of the
complex. Converting the Chern character (the sum) into the Euler
character (the product) we find (\ref{eq:1-loop}). 

For the Dirac complex associated with the hypermultiplet of mass $m$,  the weight
space $w$ contributes as 
\begin{equation}
\frac { t_1^{\frac 1 2}  t_2^{\frac 1 2}}{ (1 - t_1)(1 -t_2)} \omega \mu
\end{equation}
where $ \omega = e^{ i w \cdot \alpha}, \mu = e^{ i m }$.  This can be
seen from Atiyah-Singer formula for the Dirac complex $ S^{+}
\stackrel{D}{\to} S^{-}$ with numerator $
 t_1^{\frac 1 2} t_2^{\frac 1 2} +  \bar t_1^{\frac 1 2} \bar
  t_2^{\frac 1 2} -  \bar t_1^{\frac 1 2} t_2^{\frac 1 2} -
  t_1^{\frac 1 2} \bar t_2^{\frac 1 2}$ or from the fact that Dirac
  complex is the twist of Dolbeault complex twisted by the square root
  of the canonical bundle. Again, expanding in positive powers of $t_1,
  t_2$ and converting the sum to the product we find the equivariant
  Euler character, or the one-loop determinant (\ref{eq:1-loop}) for the
  hypermultiplet.

The explicit expression for $Z_{\Tequiv}^{\text{inst}}(\mathbf{q})$ can be found
for example in \cite{Nekrasov:2003rj}, \cite{Nakajima_2003Lectures} and
in Y. Tachikawa's review \cite{Ta} in this volume.

\subsection{Supersymmetric configurations on  $S^4_{\ep_1,\ep_2}$}
   The path integral for the partition function of QFT with an action
   $S$  invariant under a fermionic symmetry $\delta S = 0$ localizes
   near the \emph{supersymmetric configurations},  which are the
   field configurations invariant under $\delta$.  In other words the
 supersymmetric configurations are the zeroes of the odd vector field $\delta$ in the space of all field
   configurations \cite{Witten:1988ze}. The localization theorem is the
   infinite-dimensional generalization of the Atiyah-Bott formula
   \cite{Atiyah-Bott:1984} for
   the integration of the equivariantly closed differential forms over a
   manifold on which a compact Lie group $G$ acts
   \begin{equation}
\label{eq:AtiyahBott}
     \int_M \alpha = \int_F \frac{ i_F \alpha}{ e(N_F)}
   \end{equation}
where $F \subset  M $ is the fixed point locus of $G$ action on $M$ and
$e(N_F)$ is the equivariant Euler class of the normal bundle to $F$. 

From the analysis of the equations (\ref{eq:variation}),  similar to
the $S^4$ \cite{Pestun:2007rz} and the ellipsoid case
\cite{Hama:2012bg} we expect that the only smooth field configurations
that satisfy $\delta \lambda = 0$ for the topologically trivial gauge
bundle is the trivial gauge field, vanishing scalar $\Phi^{5}$, constant
scalar $\Phi^{6} = \const = \mathring{\Phi}_6$ and a suitable auxiliary field $Y^{i}_{j}$
proportional to $\mathring{\Phi}^{6}$. It is easy to see that such
a solution exists. Under the ansatz $F_{mn}  = 0, \Phi_5 = 0$ the
equations turn into 
an overdetermined algebraic linear system of equations on $\Phi_6$ and
$Y^{i}_{j}$, and this system has one-dimensional kernel corresponding to
the zero mode of $\Phi_6$. What is more difficult to show is the absence
of other solutions, and presumably this can be shown similarly to the
analysis in \cite{Pestun:2009nn}.

With the ansatz $F_{mn} = 0, \Phi_5 = 0$ and all fermions set to zero, 
we find the explicit  supersymmetric configuration invariant under  the $\delta_{\ve,
  \eta}$ (\ref{eq:variation})
\begin{multline}
\label{eq:susyconfig}
 Y^{i}{}_{j}  = \hat Y^{i}{}_j \Phi_6,  \qquad
  \hat Y^{i}{}_{j} =  \bigg( \left( \frac {1}{ 2 f_1} - \frac {\varphi(\rho) \cos \rho  }{4 f_1} +
  \frac{  \varphi(\rho) \cos \rho }{4 f_2}\right) (\sigma_3)^{i}{}_{j}  \\ 
+ \left ( \frac{ f_3}{ 2 f_1 f_2} - \frac{ \varphi(\rho) \cos \rho  f_3}{4
    f_1 f_2} \right)  (e^{ \tfrac 1 2 (i \phi_1 + i \phi_2)} \sigma_2
 e^{ -\tfrac 1 2 (i \phi_1 + i \phi_2)})^{i}{}_{j} \bigg ) 
\end{multline}

It is straightforward to evaluate the classical action on the
supersymmetric configuration (\ref{eq:susyconfig}) and find 
\begin{equation}
  S|_{\text{susy conf}} = - \tfrac{1}{\gym^2} \tr  \mathring{\Phi}_6^2
\int_{S^4_{\ep_1,\ep_2}} \sqrt{g} d^4 x   \left(   (\tfrac 1 6 R  + M ) + 2 \hat Y^{i}{}_j \hat Y^{j}{}_{i} -
    \tfrac 1 {16} T_{\mu \nu } T^{\mu \nu} 
  \right)
\end{equation}

From the explicit solution for the $T_{\mu \nu}$
(\ref{eq:Tsolution})(\ref{eq:T-deformed})(\ref{eq:parameters}), the
$\hat Y^{i}{}_{j}$ (\ref{eq:susyconfig}) and  $M$ (\ref{eq:Mscalar}) we
find that most terms in the action combine into total derivative 
\begin{multline}
\sqrt{g} ( (\tfrac 1 6 R  + M ) + 2 \hat Y^{i}{}_j \hat Y^{j}{}_{i} -
    \tfrac 1 {16} T_{\mu \nu } T^{\mu \nu} ) =\\
=f_1 f_2  r_1 r_2 \sin^3 \rho
 \sin \theta \cos \theta 
 \bigg(
\frac{ \varphi f_3  \partial_{\theta} f_2 }{2 f_1
   f_2^3}-\frac{\varphi  \partial_{\theta}f_3}{2 f_1
   f_2^2}+\frac{\varphi \partial_{\rho} f_1  \sin \rho }{2 f_1
   f_2^2} -\frac{\varphi \partial_{\rho} f_2  \sin \rho}{2
   f_2^3}\\+\frac{\varphi f_3   \tan \theta }{2 f_1
   f_2^2}-\frac{ \varphi f_3  \cot \theta }{2 f_1 f_2^2}
-\frac{\varphi \partial_{\rho}  \sin \rho }{2 f_1 f_2}
-\frac{2 \varphi  \cos \rho}{f_1
   f_2}+\frac{3}{f_1 f_2}+\frac{ \partial_\rho \varphi \sin \rho }{2
   f_2^2}+\frac{2 \varphi \cos \rho}{f_2^2} \bigg)=\\
= - \partial_{\theta} \left(r_1 r_2 \varphi  \sin 2 \theta \sin^3 \rho  
\frac{f_3}{4 f_2} \right) 
+\partial_{\rho} \left( r_1 r_2 \varphi \sin 2 \theta \sin^4 \rho 
\frac{ f_1 - f_2 }{ 4 f_2} \right) + \frac{3}{2} r_1 r_2 \sin 2 \theta \sin^3
\rho 
\end{multline}
The last term is the only term non-vanishing  after integration over
$S^4_{\ep_1,\ep_2}$. It gives 
\begin{equation}
    S|_{\text{susy conf}}  = -\frac{ 8 \pi^2 r_1 r_2}{\gym^2} \tr
    \mathring{\Phi_6}^2 
\end{equation}

Therefore, the contribution from the smooth configuration of the 
localization locus  for the partition function is\footnote{In our
  conventions $\Phi_6$ is an element of the Lie algebra of the gauge
  group. For $U(N)$ gauge group $\Phi_6$ is represented by
  anti-Hermitian matrices. The bilinear form $\la, \ra$ is the positive definite
  invariant metric on the Lie algebra normalized such
  that the length squared of the long root is  2. For $U(N)$ group  $ \tr_{\mathrm{f}}
  \Phi^2 = - \la \Phi, \Phi \ra$.}
\begin{equation}
\label{eq:X4part}
   Z_{S^4_{\ep_1,\ep_2}}^{\text{pert}} =   \int d \mathring \Phi_6  e^{ -   S|_{\text{susy conf}}}
   Z_{\text{1-loop}}(\mathring \Phi_6) = \int d \mathring \Phi_6 e^{ -
     \frac {1}{ \ep_1 \ep_2} \frac{8 \pi^2}{\gym^2} \la \mathring
   \Phi_6, \mathring \Phi_6 \ra }  Z_{\text{1-loop}}(\mathring \Phi_6)
\end{equation}
 where  $Z_{\text{1-loop}}(\Phi_6)$ needs to be computed from the
 fluctuations of the quantum fields around the supersymmetric
 background. Since mathematically such determinant is the same as a
 certain infinite-dimensional  equivariant Euler class as in the equation
 (\ref{eq:AtiyahBott}), it can be computed  \cite{Pestun:2007rz} using equivariant Atiyah-Singer index
 theorem for the transversally elliptic operators
 \cite{AtiyahCompact}. The Atiyah-Singer index theorem computes the
 index as the sum of the contributions from the fixed points: the north
 and the south pole of the $S^4_{\ep_1,\ep_2}$. The result is that the
 $Z_{\text{1-loop}}$ factorizes into the product of two factors, each
related to the one-loop factor $Z^{\text{1-loop}}_{\Tequiv}$
 (\ref{eq:1-loop}) of the gauge theory partition function in the Omega
 background coming from the north or the south pole of the $S^4_{\ep_1,\ep_2}$.
Careful application of Atiyah-Singer index theorem for the transversally
elliptic operator shows that the north pole contributes
the factor $Z^{\text{1-loop}}_{\Tequiv}$ obtained from the expansion of the
index in the positive powers of the equivariant parameters $t_1 =e^{i
  \ep_1}, t_2 = e^{ i \ep_2}$ (\ref{eq:1-loop}). The contribution of the
south pole is obtained from the expansion of the index in the negative
powers of the equivariant parameters. 

 The argument of $Z^{\text{1-loop}}_{\Tequiv}$, the equivariant parameter $\mathbf{a}$ of the gauge theory in the Omega
 background,  relates to the scalar fields on $S^4_{\ep_1,\ep_2}$ in the way
 \begin{equation}
   \mathbf{a} = v^{a} \Phi_a 
 \end{equation}
where $v^{a}$ is the vector field (\ref{eq:bilvector}). At the north
pole for the supersymmetric configuration we find 
\begin{equation}
  \mathbf{a} = -i \Phi_6 - \Phi_5 = -\ii \mathring \Phi_6
\end{equation}
On $X^4$ it is natural to assume the mass parameter pure imaginary,
since the mass can be thought as the fixed background value of the
scalar field $\Phi_6$ in the vector multiplet of gauged flavour
symmetry, so for convenience we set that mass parameters on $X^4$ are $i
\mathbf{m}$ where $\mathbf{m}$ is real. 
Then, up to an overall phase, and
assuming that the arguments $a_i$ and $m_{l \mathfrak{f}}$ in
(\ref{eq:1-loop}) are pure imaginary we  find 
\begin{equation}
  Z_{\text{1-loop}; S^4_{\ep_1,\ep_2}} = Z_{\text{1-loop}, \Tequiv}( \ii
  \mathbf{a}, \ii \mathbf{m})
\end{equation}

The classical contribution also factorizes, and using
(\ref{eq:classical}), (\ref{eq:tau}) we find the partition function (\ref{eq:X4part}) can be rewritten as 
\begin{equation}
  Z^{\text{pert}}_{S^4_{\ep_1,\ep_2}} = \int [d \mathbf{a}] |Z^{\text{pert}}_{\Tequiv}(\ep_1, \ep_2;
  \ii \mathbf{a}, \ii \mathbf{m} )|^2
\end{equation}

The above formula for the partition function takes into account only the
perturbative contribution in the localization computation around the
smooth solution of the supersymmetric equations.  However, the complete
partition function on $S^4_{\ep_1,\ep_2}$ is also contributed by the point like
instanton/anti-instanton configurations, with point instantons supported
at the north pole and the point anti-instantons supported at the south
pole~\cite{Pestun:2007rz}. This follows from the analysis of the asymptotics of the
localization equations near the north and south poles: the
supersymmetric theory on $S^4_{\ep_1,\ep_2}$ near the north pole is approximated by the gauge theory in
the Omega-background, and the supersymmetric theory on $S^4_{\ep_1,\ep_2}$ near the
south pole is approximated by the conjugated version of the gauge theory
in the Omega-background. This argument leads to the
complete formula 
\begin{equation}
\label{eq:result}
  Z_{S^4_{\ep_1,\ep_2}} = \int [d \mathbf{a}] |Z_{\Tequiv}(\ep_1, \ep_2;
  \ii \mathbf{a}; \ii \mathbf{m}; \mathbf{q})|^2
\end{equation}

\subsection{Hypermultiplets} The treatment of conformal massless hypermultiplets is straightforward
and is similar to \cite{Hama:2012bg}. The mass-term are added by
gauging the flavour symmetry, introducing the vector-multiplet for the
flavour-symmetry group and then freezing all the fields of this
flavour-symmetry vector field to zero except the constant scalar field
$(\mathring{\Phi}_0)_{\text{flavour}}$ which then plays the role of the
mass parameter.

\subsection{Open problem}
 It should be possible to classify all possible $T^2$-bundle solutions
 to generalized conformal Killing equations, construct the
 supersymmetric theories on such backgrounds and localize the partition
 function  generalizing the result (\ref{eq:result}).

\appendix
\section{Conventions and useful identities\label{se:conventions}}
\subsection{Indices\label{se:indices}}
For the 4d theories with 8 supercharges ($\mathcal{N}=2$ supersymmetry in 4d) we use the notations of the
$(0,1)$ 6d supersymmetric theories under the dimensional reduction. 
The table \ref{ta:indices}  summarizes the index notations
\begin{table}[!h]
\caption{Indices\label{ta:indices}}
  \centering
  \begin{tabular}{cccc}
   Type of indices & symbol  &  range   & fields \\
\hline
   4d space-time  vectors & $\mu, \nu, \rho, \sigma$ & $[1,\dots, 4]$ &
   gauge field $A_\mu$ \\
   $U(1)_\RS = SO(2)_\RS$ vectors  & $a,b$  &  $[5,6]$ & scalar field $\phi_a$\\
   6d vectors (the sum of above)         &  $m,n,p,q$ & $[1,\dots, 6]$   & $A_m = (A_\mu, \phi_a)$\\
   $SU(2)_{\RS}$     &  $i,j$            & $1,2$  &
   gaugino doublet $\lambda^i$;   auxiliary triplet $Y^{i}{}_{j}$
  \end{tabular}
\end{table}

The symmetrization and anti-symmetrization of tensors 
\begin{equation}
  \begin{aligned}
 t_{(m_1 \dots m_r)} &= \tfrac{1}{r!} \sum_{\sigma \in \text{Perm}(r)}
t_{m_{\sigma(1)}, \dots m_{\sigma(r)}}\\
  t_{[m_1 \dots m_r]} &= \tfrac{1}{r!} \sum_{\sigma \in \text{Perm}(r)}
  (-1)^{\sigma} t_{m_{\sigma(1)}, \dots m_{\sigma(r)}}
  \end{aligned}
\end{equation}

\subsection{Spinors}
The spinors $\lambda$ and $\ve$ in the $(0,1)$ Euclidean supersymmetric 6d
theory are the holomorphic  $SU(2)_\RS \simeq Sp(1)_\RS$
doublets of Weyl four-component spinors, of Weyl chirality $+1$, for the
6d Clifford algebra over complex numbers $\BC$.  We take $\lambda \equiv
(\lambda^i)_{i = 1,2}$ where each $\lambda^1$ and $\lambda^2$ is 6d Weyl
fermion. In total the spinor $\lambda$ has 8 complex components.\footnote{We construct the
Lagrangian and supersymmetry algebra using only holomorphic/algebraic
dependence on the spinorial components. In other words, the complex
conjugate of gaugino $(\lambda^i)$ never appears neither in the Lagrangian,
nor in the measure of the path integral, nor in the supersymmetry
transformations. The fermionic analogue of the \emph{contour of integration} in the path integral or
the \emph{reality condition} is not necessary since evaluation the Pfaffian or top
degree form is an algebraic operation.}

\subsection{Clifford algebra}
\label{se:appendix-clifford}
The $8 \times 8$ complex matrices $\gamma_m$ represent the 6d Clifford
algebra
\begin{equation}
  \{ \gamma_{m}, \gamma_{n} \} = 2 g_{mn}
\end{equation}
The chirality operator $\gamma_{*}^{\mathrm{6d}}$ anticommuting with all $\gamma_m$ is 
\begin{equation}
  \gamma_{*} = i \gamma_1 \dots \gamma_{6}; \qquad \{\gamma_{*} ,
  \gamma_m\} = 0; \qquad \gamma_*^2 = 1.
\end{equation}
The chirality of the spinors is the eigenvalue of $\gamma_*$. The
projection operators that split $\mathcal{S} = \mathcal{S}^{+}
\oplus \mathcal{S}^{-}$ are
\begin{equation}
  \gamma_{\pm} = \tfrac 1 2 (1 \pm \gamma_{*}),\qquad   \ve_{\pm} = \gamma_{\pm} \ve_{\pm} =\pm \gamma_{*} \ve_{\pm} 
\end{equation}

Explicit form of $\gamma_m$ matrices is not needed, but for
concreteness  one can recursively define the $\gamma^{(d)}_m$
matrices of size $2^{d/2} \times 2^{d/2}$  in
even dimension $d$ in terms of $\gamma^{(d-2)}_m$ as follows (see
e.g. \cite{Kennedy:1981kp})
\begin{equation}
\label{eq:stand}
  \begin{aligned}
&  \gamma^{(d)}_m   = \sigma_{3} \otimes \gamma_{m}^{(d-2)}, \quad m \in
[1, \dots, d- 2] \\
&  \gamma^{(d)}_{d-1} = \sigma_{1} \otimes 1, \qquad   \gamma^{(d)}_{d} =   \sigma_{2} \otimes 1 \\
&  \gamma^{(d)}_{*} = \sigma_3 \otimes \gamma^{(d-2)}_{*}  
  \end{aligned}
\end{equation}
where $(\sigma_0, \sigma_1,\sigma_2,\sigma_3)$ are the $2 \times 2$ Pauli matrices
\begin{equation}
\label{eq:sigma}
 (\sigma_0,\sigma_1,\sigma_2,\sigma_3) = 
(\left  (\begin{smallmatrix}
    1 & 0 \\
    0 & 1 
  \end{smallmatrix}),
 (\begin{smallmatrix}
    0 & 1 \\
    1 & 0 
  \end{smallmatrix}),
(  \begin{smallmatrix}
    0 & -i \\
    i  &  0 
  \end{smallmatrix}),
(  \begin{smallmatrix}
    1 &  0 \\
    0 & -1 
  \end{smallmatrix}) \right).
\end{equation}
We use antisymmetric multi-index notations 
\begin{equation}
  \gamma_{m_1 \dots m_r}  = \gamma_{[m_1} \dots \gamma_{m_r]}.
\end{equation}
and we use underline notation for the multi-index
\begin{equation}
\gamma_{\underline{r}} \quad \text{is one of} \quad  \gamma_{m_1 \dots m_r} 
\end{equation}
In the contraction of multi-index we use  non-repetitive summation 
\begin{equation}
  A^{\underline{p}}{} B_{\underline{p}} \equiv \sum_{m_{1} < \dots <
    m_{p}} A^{m_1 \dots m_p} B_{m_1 \dots m_p} = \frac{1}{p!}A^{m_1 \dots m_p} B_{m_1 \dots m_p}
\end{equation}
For the forms we use  component and slashed notation 
\begin{align}
\label{eq:slashed}
& \omega \equiv \tfrac{1}{r!} \omega_{\mu_1 \dots \mu_r} dx^{\mu_1} \wedge
 \dots  \wedge dx^{\mu_r}  \qquad 
& \slashed{\omega} = \omega_{\mu_1 \dots \mu_r} \gamma^{\mu_1 \dots
  \mu_r} 
\end{align}
Contraction identity
\begin{equation}
    \gamma^{m} \gamma_{\underline{r}} \gamma_m = (-1)^{r} (d - 2r)
    \gamma_{\underline{r}}
\end{equation}
Multi-index contraction identity
\begin{equation}
  \gamma^{\underline{p}} \gamma_{\underline{r}}\gamma_{\underline{p}} = 
\Delta(d,r,p) \gamma_{\underline{r}}
\end{equation}
with 
\begin{equation}
  \Delta(d,r,p)=  (-1)^{p(p-1)/2} (-1)^{rp} \sum_{q=\mathrm{max}(p+r-d,0)}^{\mathrm{min}(r,p)}
   (-1)^{q} \binom{r}{q} \binom{d-r}{p-q}
\end{equation}
The contraction formula and the completeness of
$(\gamma_{\underline{p}})_{p \in [0,\dots, d]}$ for $d \in 2 \BZ$ in the
  matrix algebra of $2 ^{d/2} \times 2^{d/2}$ matrices 
 implies the Fierz identity 
  \begin{equation}
\label{eq:Fierzgeneral}
    (\gamma^{\underline{r}}) ^{\alpha_1}_{\,\,\alpha_2}
    (\gamma_{\underline{r}})^{\alpha_3}_{\,\,\,\alpha_4} =
    \sum_{k=0}^{d} \tilde \Delta(d,r,k) 
(\gamma^{\underline{k}})^{\alpha_1}_{\,\,\ \alpha_4}
(\gamma_{\underline{k}})^{\alpha_3}_{\,\,\, \alpha_2}
  \end{equation}
where
 \begin{equation}
   \tilde \Delta (d,r,k) =  (-1)^{ \frac{k(k-1)}{2}} 2^{- \frac d 2} \Delta(d,r,k)
 \end{equation}

The terms with  $k > \frac d 2$ in the Fierz identity 
are conveniently represented as 
  \begin{equation}
    \gamma_{\underline{k}} \gamma_{*} = (-1)^{r(r-1)/2} i^{-n/2} \gamma_{\underline{k}^\vee}
  \end{equation}
where $\gamma_{\underline{k}^{\vee}}$ is complementary in indices of 
  $\gamma_{\underline{k}}$ with a proper permutation sign. 
The Fierz identity is 
\begin{equation}
\label{eq:chiralFierz}
      (\gamma^{\underline{l}}) ^{\alpha_1}_{\,\,\alpha_2}
    (\gamma_{\underline{l}})^{\alpha_3}_{\,\,\,\alpha_4} =
    \sum_{k=0}^{d/2} \tilde \Delta(d,l,k) 
(\gamma^{\underline{k}})^{\alpha_1}_{\,\,\ \alpha_4}
(\gamma_{\underline{k}})^{\alpha_3}_{\,\,\, \alpha_2}
+   (-1)^{d/2}  \sum_{k=0}^{d/2 - 1} \tilde \Delta(d,l,d- k) 
(\gamma^{\underline{k}} \gamma_*)^{\alpha_1}_{\,\,\ \alpha_4}
(\gamma_{\underline{k}} \gamma_*)^{\alpha_3}_{\,\,\, \alpha_2}.
\end{equation}
This form is useful when applied to the chiral spinors. 

\subsection{Spinor bilinears}

The spinor representation space $\mathcal{S}$ can 
be equipped with an invariant complex bilinear form  $(, ): \mathcal{S}
\otimes \mathcal{S} \to \BC$. 
In components we write
\footnote{Often in the physics literature the dual spinor $\eta_{\beta} = \eta^{\alpha}C_{\alpha \beta}$ (an element of the dual space $\mathcal{S}^{\vee}$) 
is denoted $\bar \ve$ and is called \emph{Majorana} conjugate to
$\ve$. We have chosen here to avoid the bar notation to avoid 
confusion with complex conjugation.}
\begin{equation}
 (\conj{\eta}   \ve) := \eta^{\alpha} C_{\alpha \beta} \ve^{\beta}
\end{equation}
where $C$ is a matrix representing the bilinear form.

All operators $ \gamma_{\underline{r}} $ are symmetric or
antisymmetric with respect to $C$.   The symmetry of
 $C  \gamma_{\underline{r}}$ depends on the dimension $d$ and is summarized in the table \ref{table:symmetries}
\begin{table}[!h]
\label{table:symmetries}
\caption{Symmetries of $C\gamma_{\underline{r}}$}
\begin{tabular}[center]{r|c|c|c|c}
$d \bmod 8 $      &    2      &       4        &        6      &   8 \\
\hline
   $C_1$     & $++--$    &   $  -++- $ &    $\boxed{--++}$  & $+--+$ \\ 
   $C_2$     & $-++- $   &    $\boxed{--++} $ &    $+--+ $ &  $ ++--$ \\
\end{tabular}
\end{table}

The entries $s_0 s_1 s_2 s_3$ with $s_{r} = \pm 1$ denote the transposition symmetry 
of $C\gamma_{\underline{r}}$ for $ r \bmod 4$.  There are two
choices of $C$ denoted by $C_1$ and $C_2$ in the table. 
In representation (\ref{eq:stand}) one can take
  \begin{equation}
    \begin{aligned}
    C_1 = \cdots \otimes \sigma_1 \otimes \sigma_2 \otimes \sigma_1 \\
    C_2 = \cdots \otimes \sigma_2 \otimes \sigma_1 \otimes \sigma_2.
    \end{aligned}
 \end{equation}
The bilinear form $C_2$ for spinors in even dimension $d$ can be also used 
as the bilinear form for spinors in $d+1$-dimensions. 
For the theories with $8$ supercharges in $d=4,5,6$ dimensions
we are using  $C$ highlighted  in the table \ref{table:symmetries}.

The matrices $C\gamma_{\underline{r}}$ represent bilinear forms on
$\mathcal{S}$ valued in $r$-forms, in other words, for spinors $\eta$
and $\ve$ the 
\begin{equation}
  \omega_{\underline{r}}= (\conj{\eta} \gamma_{\underline{r}} \ve)
\end{equation}
 transform covariantly as the rank $r$ form.  Since
\begin{equation}
  \gamma_{*} C = (-1)^{\tfrac d 2 } C \gamma_*
\end{equation}
it follows that\footnote{Consistent with the fact that for $\tfrac  d 2
  \in 2\BZ$ 
the tensor product $\mathcal{S}_{+} \otimes
  \mathcal{S}_{-}$ contains odd rank forms;
  and $ \mathcal{S}_{+} \otimes  \mathcal{S}_{+}, \mathcal{S}_{-} \otimes \mathcal{S}_{-}$ contains even rank
  forms; in particular for $\tfrac d 2
  \in 2 \BZ$ the representation $\mathcal{S}^{\pm}$ is dual to $\mathcal{S}^{\pm}$; while  for
  $\tfrac d 2 \in 2 \BZ + 1$ the representation $\mathcal{S}^{\pm}$ is dual to
  $\mathcal{S}^{\mp}$. }
\begin{equation}
  \begin{aligned}
  (\conj{\eta}  \gamma_{\underline{r}} \ep) &=( \conj{\eta_{+}}
  \gamma_{\underline{r}} \ep_{+})
 + (\conj{\eta_{-}}  \gamma_{\underline{r}} \ep_{-}), \qquad  \tfrac
  d 2 + r   \in 2
  \BZ  \\
( \conj{ \eta} \gamma_{\underline{r}} \ep) &= (\conj{\eta_{-}}  \gamma_{\underline{r}}
  \ep_{+}) + (\conj{\eta_{+}}  \gamma_{\underline{r}} \ep_{-}),  \qquad \tfrac d 2
  + r  \in 2 \BZ + 1
  \end{aligned}
\end{equation}

In $d=6$ the bilinears in the spinors of the same chirality 
transform as forms of odd rank; 
while the bilinears in the spinors of opposite chirality transform as
forms of even rank. 
\begin{equation}
d=6: \qquad 
  \begin{cases}
  (\conj{\ve_+} \gamma_{\underline{r}} \ve_+') \neq 0 \quad \text{only for $r \in \{1,3,5\}$}\\
( \conj{\eta_{-}}  \gamma_{\underline{r}} \ve_{+}) \neq 0 \quad \text{only for $r \in \{0,2,4,6\}$}
  \end{cases}
\end{equation}
The bilinear form valued in $1$-forms is antisymmetric in $d=6$ for either choice of $C$. To construct the standard
fermionic  action $(\lambda \gamma^m D_m \lambda)$ we need the symmetric 
$1$-form valued bilinear form. For the minimal  6d $(0,1)$ supersymmetry
we introduce a $SU(2)_{\RS}$-doublet of Weyl fermions $(\lambda^{i})_{i=1,2}$ and then
use  $C \otimes \ep$, where $\ep = \ep_{ij}$ is the standard $2\times2$
antisymmetric symbol,  as the symmetric bilinear form on the
$\mathcal{S^{+}}\otimes \BC^{2}$. The resulting 1-form valued bilinear 
is symmetric and there is a proper fermionic kinetic action 
\begin{equation}
  (\lambda^{i} \slashed{D} \lambda_i) \equiv  (\lambda^{i}
 \slashed{D} \lambda^j) \ep_{ji}
\end{equation}

We use the standard antisymmetric $2 \times 2$
tensor $\ep_{ij}$ to raise and lower the $SU(2)_\RS$ indices $i,j$
in  the pattern $^i{}_i$:
\begin{equation}
\label{eq:ij}
  \begin{aligned}
&  \lambda^{i} := \ep^{ij} \lambda_{j}, \qquad   \lambda_{j} := \lambda^{i} \ep_{ij}\\
&  \ep^{ij} \ep_{ik} = \delta^{j}_{k}, \qquad   ( \conj{\ve}^{[j} \eta^{i]}) = \tfrac 1 2 \ep^{ij} (\conj{\ve}^k \eta_k)
  \end{aligned}
\end{equation}
Whenthe  $SU(2)_\RS$ indices are omitted, the contraction  $^i{}_i$ is
assumed
\begin{equation}
(\conj{\ve} \gamma_{\underline{r}} \ve')  \equiv  ( \conj{\ve}^{i} \gamma_{\underline{r}} \ve'_i) 
\end{equation}

\subsection{$d=6$ Fierz identities}
For $d=6$ and  $l=1$ we find 
  \begin{table}[!h]
    \centering
    \begin{tabular}{c|ccccccc}
$k$    &  0 &  1 & 2 & 3 & 4 & 5 & 6\\
\hline
$\tilde \Delta(6,1,k)$ & $ \tfrac 3 4 $  & $-\tfrac{1}{2}$ & $-\tfrac {1}{4}$  & $0$ & $
-\frac{1}{4} $ & $\frac{1}{2}$ & $\tfrac{3}{4}$ 
    \end{tabular}
  \end{table}

Notice that $\tilde \Delta(6,1,k) = (-1)^k \tilde
\Delta(6,1,6-k)$. Therefore if we project Fierz identity (\ref{eq:chiralFierz})
with $\gamma_{+}$ applied to the $\alpha_2$ 
and $\alpha_4$ indices, we find that terms with even $k$ vanish. In
addition the middle term $k = 3$ vanishes too. Finally
\begin{equation}
\label{eq:6dFierz1}
\boxed{         (\gamma^{\underline{1}}) ^{\alpha_1}_{\,\,\alpha_2}
    (\gamma_{\underline{1}})_{\alpha_3 \alpha_4} =
-  (\gamma^{\underline{1}})^{\alpha_1}_{\,\,\ \alpha_4}
(\gamma_{\underline{1}})_{ \alpha_3  \alpha_2}   \quad \quad
\text{projected by $(\gamma_{\pm})^{\alpha_2}_{\,\,\, \alpha_2'} (\gamma_{\pm})^{\alpha_4}_{\,\,\, \alpha_4'}$}  }
\end{equation}
A frequently used form of the above identity involves
cylic permutation of three $+$-chiral spinor doublets
$\ve^i, \kappa^i, \lambda^i$. Taking the sum of  
\begin{equation}
  \begin{aligned}
  (\ve^j \gamma_{m} \kappa_j) \gamma^m \lambda^i = - (\ve^j \gamma_m
  \lambda^i) \gamma^m \kappa_j \\
 (\lambda^j \gamma_m
  \kappa_j) \gamma^m \ve^i = -  (\lambda^j \gamma_{m} \ve^i) \gamma^m \kappa_j 
  \end{aligned}
\end{equation}
we find 
\begin{equation}
\label{eq:Fierz3}
\boxed{    (\ve \gamma_{m} \kappa) \gamma^m \lambda +  (\lambda \gamma_m
  \kappa) \gamma^m \ve +  (\ve \gamma_m
  \lambda) \gamma^m \kappa = 0}
\end{equation}

Now we consider projection of 6d Fierz identity at $p=1$ on spinors of
opposite chirality. Take $+$-chiral doublet $\ve^i$ and $-$-chiral
doublet  $\eta^i$. We find 
  \begin{equation}
    \begin{aligned}
\label{eq:2}
&    (\conj{\ve^{j}} \gamma^{\underline{1}} \ve_j) \gamma_{\underline{1}} \eta^i = 
\tfrac 3 2 (\conj{\ve^{j}} \eta^i) \ve_j  - \tfrac 1 2 (\conj{\ve^{j}}
\gamma^{\underline{2}} \eta^i) \gamma_{\underline{2}} \ve_j \\
& (\conj{\ve^{j}}
\gamma^{\underline{2}} \eta^i) \gamma_{\underline{2}} \ve_j =
-\tfrac{5}{4}  (\conj{\ve^{j}} \gamma^{\underline{1}} \ve_j) \gamma_{\underline{1}} \eta^i
\end{aligned}
\end{equation}
where at $l=2$ the explicit coefficients in (\ref{eq:chiralFierz}) are
given as follows: 
 \begin{table}[!h]
    \centering
    \begin{tabular}{c|ccccccc}
$k$    &  0 &  1 & 2 & 3 & 4 & 5 & 6\\
\hline
$\tilde \Delta(6,2,k)$ & $ -\tfrac {15} 8 $  & $-\tfrac{5}{8}$ & $-\tfrac{1}8$  & $-\tfrac{3}{8}$ & $
+\tfrac{1}{8} $ & $-\tfrac{5}{8}$ & $+\tfrac{15}{8}$ 
    \end{tabular}
  \end{table}

Hence, from the equations (\ref{eq:2}) we find  another useful 6d Fierz identity
\begin{equation}
\label{eq:6dFierz2}
\boxed{   (\conj{\ve^{j}} \gamma^{m} \ve_j) \gamma_m   \eta^i = 4
   (\conj{\ve^{j}} \eta^i) \ve_j, \qquad (\ve = \gamma_{+} \ve, \quad
   \eta = \gamma_{-} \eta)}
\end{equation}

\subsection{6d $(0,1)$ theory conventions}
The spinor $\ve$ is $+$-chiral, the spinor $\eta$ is $-$ chiral 
\begin{align}
&\ve  =   \gamma_* \ve = \gamma_{+} \ve, \qquad &\eta =   - \gamma_{*}\eta = \gamma_{-} \eta 
\end{align}

The  tensor field $T_{\mu \nu a}$ is 6d anti-self-dual,  $*_{6d}
T = -T$. Useful contraction identities 
\begin{align}
\label{eq:Tzero}
& \gamma^{\mu} \slashed{F} = \slashed{F} \gamma^{\mu} - 4 F_{m \mu}
\gamma^{m}, \qquad  \slashed{T} = T_{\mu \nu a} \gamma^{\mu \nu a} \\
& \gamma^\mu \slashed{T} \gamma_{\mu} = \gamma^a \slashed{T} \gamma_{a} =  
 \gamma^m \slashed{T} \gamma_{m}   = 0\\
&T_{\mu \nu a} \gamma^{\mu\nu} = \tfrac 1 2 \{ \slashed{T}, \gamma_a\}, \qquad
T_{\mu \nu b} \gamma^{\nu b}  = \tfrac 1 4 \{ \slashed{T}, \gamma_{\mu} \}
\label{eq:Tamove}
\end{align}

The Bianchi identity on the field strength
\begin{equation}
\label{eq:Bianchi}
  \begin{aligned}
    &  D_{m} F_{pq}  + D_{q} F_{mp} + D_{p} F_{qm} = 0, \qquad    \gamma^{mpq} D_{m} F_{pq} = 0
  \end{aligned}
\end{equation}

Positive chirality of  $\ve \equiv \ve^{+}$ and negative chirality of $T
\equiv T^{-}$ implies
\begin{equation}
\label{eq:Tep}
\begin{aligned}
& \slashed{T} \gamma_{\underline{2r}}\ve = 0\\
&  (\conj{\ve} \gamma_{\underline{r}} \ve) = 0, \quad r \bmod 4  \in
\{2,3\}; \qquad 
  (\conj{\ve^{(i}} \gamma_{\underline{r}} \ve^{j)}) = 0, \quad r \bmod 4 \in
\{0,1\} \\
& \gamma_{\rho} \{ \slashed{T}, \gamma_{a}\}\ve = \{ \slashed{T},
\gamma_\rho \} \gamma_a \ve , \qquad  \tfrac{1}{2} T_{\mu \nu a} \gamma_{\rho} \gamma^{\mu \nu} \ve = T_{\rho
  \nu b} \gamma^{\nu b} \gamma_{a} \\
&  \gamma^{\rho} T_{\mu \nu a} \gamma^{\nu a} \ve = T_{\rho \nu a} \gamma^{\nu
   a} \gamma_{\mu} \ve, \qquad  \gamma_a T_{\mu \nu b} \gamma^{\mu \nu} \ve = T_{\mu \nu a} \gamma^{\mu
  \nu} \gamma_b \ve\\
&  T_{\mu \nu a} T_{\rho \sigma b} \gamma^{\rho} \gamma^{\mu \nu}
\gamma^{\sigma b} \ve = 4 T_{\rho \nu a} \gamma^{\nu}T_{\rho \sigma b} 
\gamma^{\sigma b} \ve = 4 T_{\rho \nu a} T_{\rho \nu b} \gamma^{b} \ve 
\end{aligned}
\end{equation}

The spin-connection and the metric curvatures 
\begin{equation}
\label{eq:curvatures}
\begin{aligned}
&  D_{\mu} v^{\hat \rho} =     \partial_\mu v^{\hat \rho} + \omega^{\hat
  \rho}{}_{\hat \sigma \mu} v^{\hat \sigma} \\
&  R^{\hat \rho}{}_{\hat \sigma \mu \nu} = [D_{\mu}, D_{\nu}]^{\hat
    \rho}_{\hat \sigma}, \qquad 
  R_{\sigma \nu} =   R^{\mu}{}_{\sigma \mu \nu}, \qquad 
 R = R^{\mu}{}_{\mu} 
\end{aligned}
\end{equation}

The covariant derivative on spinors, the curvature and the  Lichnerowicz formula
\begin{equation}
\begin{aligned}
D_{\mu} \ve^i = \partial_\mu \ve^i + \tfrac 1 4 \omega^{ \rho 
  \sigma}{}_{\mu}\gamma_{ \rho  \sigma} \ve^i + (V^{\RS}_\mu)^{i}{}_{j} \ve^j \\
\label{eq:Lichner}
 \slashed{D}^2  = D^{\mu} D_{\mu}  - \tfrac {1}{4} R + \tfrac 1 2
 \slashed{F}^\RS_{V}
\end{aligned}
\end{equation}
where $ (V^{\RS}_\mu)^{i}{}_{j} $ is the $SU(2)_{\RS}$-connection.

\subsection{Supersymmetry equations}
The divergence of the first equation in the system (\ref{eq:conformalK}) implies
\begin{equation}
\label{eq:firstdiv}
   D^\mu D_\mu \ve - \tfrac{1}{16} [D^\mu \slashed{T}] \gamma_\mu \ve 
- \tfrac 1 4 \slashed{T} \eta 
=
 \tfrac 1 4 \slashed{D}^2 \ve 
\end{equation}
which together with Lichnerowicz formula (\ref{eq:Lichner}) produces 
\begin{equation}
\label{eq:1}
  \tfrac{1}{ 4} \slashed{D}^2 \ve+ \tfrac{1}{3} ( \tfrac 1 4 R \ve  - \tfrac{1}{2}
  \slashed{F}_\RS \ve 
  - \tfrac{1}{16} [D^{\mu} \slashed{T}] \gamma_\mu \ve - \tfrac 1 4
  \slashed{T} \eta ) = 0
\end{equation}
and the linear combination with the second equation in
(\ref{eq:conformalK}) produces 
\begin{equation}
  \tfrac 1 4 \slashed{D}^2 \ve  = -\tfrac{1}{2}(
 \tfrac 1 6 R + M) \ve +
  \tfrac{1}{16} [D^{\mu} \slashed{T}] \gamma_\mu \ve 
\end{equation}

\subsection{The 6d and 4d spinor conventions}
As in (\ref{eq:stand}) we take
\begin{equation}
\begin{aligned}
\label{eq:gamma6-new}
  \gamma_\mu^{(6)} =
 ( \begin{smallmatrix}
    \gamma_\mu^{(4)} & 0 \\
    0             & - \gamma_{\mu}^{(4)}
  \end{smallmatrix}),\quad
  \gamma_5^{(6)} =
 ( \begin{smallmatrix}
  0  & 1  \\
  1 & 0 
  \end{smallmatrix}),\quad
  \gamma_6^{(6)} =
 ( \begin{smallmatrix}
    0      &-i \\
   i        &  0
  \end{smallmatrix})\\
\quad
  C^{(6)} =
(  \begin{smallmatrix}
    0 &  - c^{(4)}\gamma_{*}^{(4)}  \\
   - c^{(4)} \gamma_{*}^{(4)}   & 0 
  \end{smallmatrix}),\quad 
  \ve^{(6)}_{+}    =
  (\begin{smallmatrix}
    \ve^{(4)}_{+} \\
    \ve^{(4)}_{-}
  \end{smallmatrix}),\quad 
  \eta^{(6)}_{-}    =
  (\begin{smallmatrix}
    \eta^{(4)}_{-} \\
  -  \eta^{(4)}_{+}
  \end{smallmatrix})
\end{aligned}
\end{equation}
where $\ve_{\pm}^{(4)}$ denote the $\pm$-chiral spinors of the 4d
Clifford algebra with respect to $\gamma_{*}^{(4)}$, the $C^{(6)}$ is the bilinear
form for the 6d Clifford algebra  of type $(--++)$,  and $c^{(4)}$ is the bilinear form of
the 4d Clifford algebra of type $(--++)$. 
In these conventions the bilinears computed in 4d and 6d notations
agree:
\begin{equation}
  \begin{aligned}
    \ve_{+}^{(6)} C^{(6)} \eta_{-}^{(6)}=   \ve_{+}^{(4)} c^{(4)} \eta_{+}^{(4)} + \ve_{-}^{(4)} c^{(4)} \eta_{-}^{(4)}\\
    \ve_{+}^{(6)} C^{(6)}  \gamma_{\mu}^{(6)} \tilde \ve_{+}^{(6)}=
    \ve_{+}^{(4)} c^{(4)} \gamma_{\mu}^{(4)}  \tilde \ve_{-}^{(4)} +
    \ve_{-}^{(4)} c^{(4)} \gamma_{\mu}^{(4)}\tilde \ve_{+}^{(4)}
  \end{aligned}
\end{equation}
For the explicit form of spinors we use 4d gamma-matrices, the 4d
chirality operator and 4d bilinear form in terms of (\ref{eq:sigma})
\begin{equation}
\label{eq:standard4d}
  \begin{aligned}
&  (\gamma_i, \gamma_4)  = (\sigma_2 \otimes \sigma_i, \sigma_1 \otimes
  \sigma_0 ), \\
&\gamma_{*}^{(4)} =  -\gamma_1 \dots \gamma_4 = -\sigma_3 \otimes \sigma_0\\
& c^{(4)}  = -i  \sigma_0 \otimes \sigma_2 \\
  \end{aligned}
\end{equation}

We decompose
\begin{equation}
  T_{\mu \nu a} \gamma^{a} = T_{\mu \nu -} \gamma^{-} + T_{\mu \nu +} \gamma^{+}
\end{equation}
in terms of 
\begin{equation}
  \begin{aligned}
&  T_{\mu \nu -}     = (T_{\mu \nu 5} - i T_{\mu \nu 6} ) \qquad
  \gamma^{-} =\tfrac 1 2 ( \gamma^5 + i \gamma^6)  \qquad \gamma^- \gamma^{56}_* = -
  \gamma^{-} \\
&  T_{\mu \nu +}     = (T_{\mu \nu 5} + i T_{\mu \nu 6} ) \qquad
  \gamma^{+} = \tfrac 1 2  (\gamma^5 - i \gamma^6) \qquad \gamma^+ \gamma^{56}_* =
  + \gamma_{+} \\
  \end{aligned}
\end{equation}
with $\gamma^{56}_* = - i \gamma_{56}$.
Since $T_{\mu \nu a}$ is of negative 6d chirality, the $T_{\mu \nu \pm}$
has $\mp$ 4d chirality. We define
\begin{equation}
  T_{\mu \nu}^{(4)} \equiv T_{\mu \nu +}  - T_{\mu \nu -} = 2 \ii T_{\mu \nu 6} 
\end{equation}
In terms of the 4d spinors  the generalized conformal Killing equation
(\ref{eq:conformalK}) takes form 
\begin{equation}
  D_{\mu} \ve  - \tfrac{1}{16} T^{(4)}_{\rho \sigma}  \gamma^{\rho \sigma} 
\gamma_\mu  \ve = \gamma_\mu \eta  \\
\end{equation}
Other 6d - 4d notational definitions are 
\begin{equation}
  \begin{aligned}
 \Phi^{\pm} &= \tfrac 1 2 (\Phi^{5} \mp i \Phi^{6}),& \qquad \Phi_{\pm} & =  (\Phi^{5} \pm i \Phi^{6})\\
    T_{\mu \nu a} \gamma^{\mu \nu} \Phi^{a} \ve_{+}^{(6)} &= T_{\mu
    \nu}^{(4)} \gamma^{\mu \nu} (\Phi^{+} \ve_{-}^{(4)} - \Phi^{-}
  \ve_{+}^{(4)}), \qquad& 
 \Phi^a \gamma_a \eta &= 2 \Phi^{-} \eta_{-}^{(4)} - 2 \Phi^{+} \eta_{+}^{(4)}\\
      \end{aligned}
\end{equation}

\section{Supersymmetry algebra}

\subsection{The off-shell closure of the supersymmetry on the vector multiplet}
Here we explicitly compute 
$\delta^2$ on vector multiplet for $\delta$ defined by:
\begin{equation}
  \begin{aligned}
&  \delta A_{m} = \tfrac  1 2 \conj{\lambda^i}  \gamma_m \ve_i \\
&  \delta \lambda^i = - \tfrac 1 4 F_{mn} \gamma^{mn} \ve^i  + 
Y^{i}_{\,\,\, j} \ve^{j} + \Phi_a \gamma^{a} \eta^i+   \tfrac 1 8 T_{\mu \nu a} \Phi^a \gamma^{\mu  \nu} \ve^i \\
&  \delta Y^{ij} = - \tfrac 1 2 (\conj{\ve^{(i} } \slashed{D}
\lambda^{j)}) \\
  \end{aligned}
\end{equation}
provided that spinors $(\ve, \eta)$ with $\eta  = \tfrac 1 4 \slashed{D} \ve$ satisfy
generalized conformal Killing equations (\ref{eq:conformal2}). 

We find a contribution of
several terms in $\delta_{}^2 \lambda^i$. In  the flat space we  drop the terms
proportional to $D_{\mu} \ve$, $\eta$ and $T$  and find 
\begin{equation}
  \delta_{\text{flat}}^2 \lambda^i = \delta (  - \tfrac 1 4 F_{mn} \gamma^{mn} \ve^i) +
\delta( Y^{i}_{\,\,\, j} \ve^{j}) 
\end{equation}
with 
\begin{multline}
\begin{aligned}
&  \delta (  - \tfrac 1 4 F_{mn} \gamma^{mn} \ve^i)= -\tfrac 1 4 D_p (\conj{\lambda^j} \gamma_q \ve_j)
  \gamma^{pq} \ve^i = \tfrac 1 4 (\conj{\ve} \slashed{D} \lambda) \ve^i -
  \tfrac 1 4 (\conj{\ve^j} \gamma_q D_p \lambda_j) \gamma^p \gamma^q \ve^i  \stackrel{(\ref{eq:6dFierz1})}{=}  \\
&\qquad = \tfrac 14 (\conj{\ve} \slashed{D} \lambda) \ve^i + 
\tfrac 1 4 (\conj{\ve} \gamma^q \ve) D_q \lambda^i  - \tfrac 1 8 (\conj{\ve} \gamma_q
\ve) \gamma^q \slashed{D} \lambda^i \stackrel{(\ref{eq:6dFierz2})}{=} \tfrac 1 4 (\conj{\ve} \gamma^q \ve) D_q \lambda^i 
 + \tfrac 14 (\conj{\ve} \slashed{D} \lambda) \ve^i - \tfrac 1 2
 (\conj{\ve^j} \slashed{D} \lambda^{i}) \ve_j 
\end{aligned}
\end{multline}
and together with the $\delta( Y^{i}_{\,\,\, j} \ve^{j}) $ we find
\begin{equation}
  \begin{aligned}
& \delta^2_{\text{flat}} \lambda^i  =  \tfrac 1 4 (\conj{\ve} \gamma^q \ve) D_q \lambda^i 
 + \tfrac 14 (\conj{\ve^j} \slashed{D} \lambda_j) \ve^i - \tfrac 1 2
 (\conj{\ve^{[j}} \slashed{D} \lambda^{i]}) \ve_j
 \stackrel{(\ref{eq:ij})}{=}  \tfrac 1 4 (\conj{\ve} \gamma^q \ve) D_q \lambda^i 
  \end{aligned}
\end{equation}
Next we  account for $D_{\mu} \ve$ and $\eta$ terms, still
keeping  $T=0$. The transformation would be complete on conformally flat
space. 
 The $\delta^2_{\text{cflat}} \lambda $ acquires new contributions 
\begin{equation}
  \delta^2_{\text{cflat}}  \lambda^i = \delta^2_{\text{flat}} \lambda^i  + \mathbf{term_c} 
\end{equation}
where
\begin{equation}
\label{eq:termc}
  \begin{aligned}
& \mathbf{term_c} =  - \tfrac 1 4 (\conj{\lambda} \gamma_q
 \gamma_\mu  \eta) \gamma^{\mu q} \ve^i + \tfrac 1 2 (\conj \lambda \gamma_a
   \ve) \gamma^a \eta^i \stackrel{ (\ref{eq:Fierz3})\,\mathrm{on}\,\,\gamma_q{}\gamma^q}{=} \\
& \qquad = -\tfrac 1 2 (\conj{\ve} \gamma_q \lambda) \gamma^q
  \eta + \tfrac 1 4 (\conj{\eta} \gamma_{\mu q} \ve) \gamma^{\mu q}
  \lambda + (\conj{\eta} \ve) \lambda + (\conj \eta \lambda) \ve 
  \end{aligned}
\end{equation}
Then we expand  the middle term in 4d indices 
\begin{equation}
   \tfrac 1 4 (\conj{\eta} \gamma_{\mu q} \ve) \gamma^{\mu q}    \lambda
   =   
+  \tfrac 1 8 (\conj{\eta} \gamma_{\mu \nu} \ve) \gamma^{\mu \nu}  \lambda    
+  \tfrac 1 8 (\conj{\eta} \gamma_{p q} \ve) \gamma^{p q}  \lambda 
-    \tfrac 1 8 (\conj{\eta} \gamma_{ab} \ve) \gamma^{ab}   \lambda 
\end{equation}
and apply Fierz identity (\ref{eq:Fierzgeneral}) to the first and the last term in (\ref{eq:termc}) to
find 
\begin{equation}
  \begin{aligned}
  -\tfrac 1 2 (\conj{\ve}^j \gamma_q \lambda_j) \gamma^q
  \eta^i &= - \tfrac 3 4 (\conj{\ve}^j \eta^i) \lambda_j + \tfrac 1 8
  (\conj{\ve}^j \gamma_{pq} \eta^{j}) \gamma^{pq} \lambda_j \\
(\conj{\eta}^j \lambda_j) \ve^i &= \tfrac 1 4 (\conj{\eta}^{j} \ve^{i}) \lambda_j -
\tfrac 1 8 (\conj{\eta}^{j} \gamma_{pq} \ve^{i}) \gamma^{pq} \lambda_j
  \end{aligned}
\end{equation}
All $\gamma_{pq} \gamma^{pq}$ terms are cancelled using
(\ref{eq:ij}) and the scalar terms are simplified as
\begin{equation}
  (\conj{\eta} \ve) \lambda -  \tfrac 3 4 (\conj{\ve}^j \eta^i)
  \lambda_j+ \tfrac 1 4 (\conj{\eta}^{j} \ve^{i}) \lambda_j = \tfrac
  {3}{ 4} (\conj{\eta} \ve) \lambda + (\conj{\eta}^{(i} \ve^{j)})
  \lambda_j 
\end{equation}
and the contribution from the non-flat but conformally flat terms is 
\begin{equation}
  \mathbf{term_c} = 
+  \tfrac 1 8 (\conj{\eta} \gamma_{\mu \nu} \ve) \gamma^{\mu \nu}  \lambda    
-    \tfrac 1 8 (\conj{\eta} \gamma_{ab} \ve) \gamma^{ab}   \lambda +
\tfrac
  {3}{ 4} (\conj{\eta} \ve) \lambda + (\conj{\eta}^{(i} \ve^{j)})
  \lambda_j
\end{equation}
Then we compute the  $T$-terms in
\begin{equation}
    \delta^2 \lambda^i = \delta^2_{\text{cflat}} \lambda^i + \mathbf{term_T}
\end{equation}
and find 
\begin{equation}
  \begin{aligned}
&  \mathbf{term_T} =  \tfrac{1}{64} (\conj{\ve} \gamma_\mu \slashed{T} \gamma_{q}
  \lambda) \gamma^{\mu q} \ve + \tfrac{1}{16} T_{\mu \nu a} (\conj{\ve} \gamma^{a}
  \lambda)
  \gamma^{\mu \nu} \ve  \stackrel{~(\ref{eq:Fierz3})\,\,\mathrm{on }\gamma_q\gamma^q, (\ref{eq:Tamove})}{=} \\ 
& \qquad =  -\tfrac{1}{64} (\conj{\ve} \gamma_\mu \slashed{T}
    \gamma_q \ve) \gamma^{\mu q} \lambda + \tfrac{1}{64} (\conj{\ve} \gamma_q \lambda) \gamma^{\mu}
    \gamma^q \slashed{T} \gamma_\mu \ve + \tfrac{1}{32}  (\conj{\ve} \gamma_{a}
  \lambda)
 \slashed{T} \gamma^{a}  \ve =\\
& \qquad =  -\tfrac{1}{64} (\conj{\ve} \gamma_\mu \slashed{T}
    \gamma_q \ve) \gamma^{\mu q} \lambda   + \tfrac{1}{32} (\ve \gamma_p \lambda)  \slashed{T}
    \gamma^p \ve    \stackrel{~(\ref{eq:Fierz3})\,\,\mathrm{on }\gamma_p\gamma^p}{=}  \\
& \qquad =  -\tfrac{1}{64} (\conj{\ve} \gamma_\mu \slashed{T}
    \gamma_q \ve) \gamma^{\mu q} \lambda - \tfrac{1}{64} ( \conj{\ve}
    \gamma_q \ve) \slashed{T} \gamma^q \lambda
    \stackrel{~(\ref{eq:Tamove})} = \tfrac{1}{32} T_{\mu \nu  a} (\conj{\ve} \gamma^a \ve)
  \gamma^{\mu \nu }  \lambda
  \end{aligned}
\end{equation}
The $\mathbf{term_T}$ can be combined  with the $\mathbf{term_c}$:
\begin{equation}
  \tfrac{1}{8} (\eta \gamma_{\mu \nu} \ve) \gamma^{\mu \nu} \lambda 
+ \tfrac{1}{32} T_{\mu \nu  a} (\conj{\ve} \gamma^a \ve)
  \gamma^{\mu \nu }  \lambda = 
\tfrac{1}{16} D_{\mu} (\ve \gamma_\nu \ve) \gamma^{\mu \nu} \lambda 
\end{equation}
so that finally
\begin{equation}
  \delta_{\ve,\eta}^2 \lambda^i = \tfrac 1 4 (\conj \ve \gamma^m \ve) D_m \lambda^i + 
\tfrac{1}{16} D_{\mu} (\ve \gamma_\nu \ve) \gamma^{\mu \nu} \lambda^i -    \tfrac 1 8 (\conj{\eta} \gamma_{ab} \ve) \gamma^{ab}   \lambda^i +
\tfrac
  {3}{ 4} (\conj{\eta} \ve) \lambda^i + (\conj{\eta}^{(i} \ve^{j)})
  \lambda_j
\end{equation}
The variation  $\delta^2_{\ve, \eta} Y^{ij}$ and $\delta^2_{\ve,\eta} A_m$ are computed
similarly. 

\subsection{The invariance of the Lagrangian}
The  4d $\mathcal{N}=2$ supersymmetric Lagrangian for vector multiplet  in curved background for vanishing
fermionic fields of Weyl multiplet is proportional to (\ref{eq:action}) 
 \begin{equation}
\tfrac 1 2 F_{mn} F^{mn}  +  \conj\lambda^i  \gamma^m D_{m}
   \lambda_i + (\tfrac{1}{6}R + M)  \Phi_a \Phi^a -  2 Y_{ij} Y^{ij} -F^{\mu \nu} T_{\mu \nu a} \Phi^a 
+ \tfrac{1}{4}  T_{\mu \nu a} T^{\mu \nu b} \Phi^a \Phi_b
 \end{equation}
The trace is implicitly implied in all terms.
To check the invariance under (\ref{eq:variation}) we first consider the
flat background with $D_{\mu} \ve = 0, \eta = 0, T=0, M =0$. After that
we will add the variational terms in conformally flat background, and
finally we will add the remaining  $T$-terms.
We find modulo total derivative 
 \begin{equation}
   \begin{aligned}
   &     \delta_{\text{flat}} (\tfrac 1 2 F^{mn} F_{mn}) = -  (\conj{\ve} \gamma^n
\lambda) D^{m} F_{mn}\\
 &\delta_{\text{flat}} (  \conj{\lambda} \gamma^m D_m \lambda) =   \tfrac 1
 2 (\conj{\lambda} \gamma^{m}  D_{m}  F_{pq} \gamma^{pq} \ve)
 + 2 Y^{i}_{\,\,\,j} (\conj{\ve^{j}} \slashed{D} \lambda_{i}) \stackrel{ (\ref{eq:Bianchi}) }{=}\\
& \qquad \qquad \qquad  = (\conj{\lambda}  \gamma^{n} \ve)D^{m} F_{mn }
  - 2 Y_{ij} (\conj{\ve^{j}} \slashed{D} \lambda^{i})\\
& \delta_{\text{flat}} (-2 Y_{ij} Y^{ij}) = 2 Y_{ij} (\ve^{i} \slashed{D} \lambda^{j})
   \end{aligned}
 \end{equation}
that all terms add to zero.
In conformally flat background the new terms appear in the variation of
fermionic kinetic term and the coupling of scalars to the curvature
\begin{equation}
\label{eq:3}
  \begin{aligned}
&    \delta_{\text{cflat}} ((\tfrac{1}{6} R+M) \Phi_a \Phi^a) =
(\tfrac{1}{6}R + M) (\conj{\lambda}
\gamma^a \Phi_a \ve)   \\
&\delta_{\text{cflat}} (  \conj{\lambda} \gamma^m D_m \lambda) = 
\delta_{\text{flat}} (  \conj{\lambda} \gamma^m D_m \lambda)  + \mathbf{term_c}
  \end{aligned}
\end{equation}
where
\begin{equation}
  \begin{aligned}
&  \mathbf{term_c}= - 2 (\conj{\lambda} [\slashed{D} \gamma^a \Phi_a \eta^i])
+ \tfrac 1 2 (\conj{\lambda} \gamma^{\mu} F_{pq} \gamma^{pq} D_{\mu}
\ve) \stackrel{~(\ref{eq:conformalK})}{=} \\
& \qquad = - 2(\conj{\lambda} \gamma^{ma} F_{ma} \eta) +2 \Phi_a  (\conj{\lambda}
\gamma^{a} \slashed{D} \eta^i) +  2  (\conj{\lambda}  F_{m a}
\gamma^{m a } \eta ) = 2 (\conj{\lambda} \gamma^a \Phi_a \slashed{D} \eta^i) \stackrel{~(\ref{eq:conformalK})}{=} \\
& \qquad = - (\tfrac{1}{6}R+M) (\conj{\lambda} \gamma^a \Phi_a \ve)
  \end{aligned}
\end{equation}
so all terms in (\ref{eq:3}) cancel when added together.

Next we consider the remaining $T$-terms for  a generic
background.
We set 
\begin{equation}
 \mathcal{ L}  = \mathcal{L}_{\text{cflat}} + \mathcal{L}_{T}
\end{equation}
where
\begin{equation}
  \mathcal{L}_{T} = -F^{\mu \nu} T_{\mu \nu a} \Phi^a 
+ \tfrac{1}{4}  T_{\mu \nu a} T^{\mu \nu b} \Phi^a \Phi_b
\end{equation}
and we find
\begin{equation}
\label{eq:5}
  \begin{aligned}
\delta(  \mathcal{L}_{T} ) =  (\conj{\lambda} \gamma^\nu
\ve) [D^{\mu} T_{\mu \nu a}] \Phi^a
+\underbrace{ {\mycolor{brown} (\conj{\lambda} \gamma^\nu
\ve) T_{\mu \nu a} F^{\mu a}}}_{\text{\textcircled{1}}}
 -\underbrace
{ \mycolor{blue} \tfrac 1 2 (\conj{\lambda} \gamma^{a} \ve) F^{\mu \nu} 
T_{\mu \nu a} }_{\text{\textcircled{2}}} +\underbrace{ \mycolor{green}\tfrac{1}{4}   (\conj{\lambda} \gamma^{a} \ve) T_{\mu \nu a} T_{\mu \nu
  b} \Phi^b}_{\text{\textcircled{3}}}
      \end{aligned}
\end{equation}
In the variation of the fermionic action the new terms are 
\begin{equation}
  \delta (  \conj{\lambda} \gamma^m D_m \lambda) =
 \delta_{\text{cflat}} (  \conj{\lambda} \gamma^m D_m \lambda) +
 \mathbf{term_{T1}} + \mathbf{ term_{T2}}
\end{equation}
where $\mathbf{term_{T1}}$ comes from $T$-terms in generalized conformal
Killing equation (\ref{eq:conformal2}) and $\mathbf{term_{T2}}$ comes
from the $T$-term in the variation $\delta_{\ve, \eta} \lambda$
(\ref{eq:variation})
\begin{equation}
\label{eq:6}
  \mathbf{term_{T1}} = 
 2  \Phi_a (\conj{\lambda} \gamma^a \slashed{D} \eta)_{T} +  \tfrac 1 {32} (\conj{\lambda}
 \gamma^{\mu} \slashed{F} \slashed{T}
\gamma_\mu  \ve) 
\end{equation}
Then we find
\begin{equation}
  \begin{aligned}
 \tfrac 1 {32} (\conj{\lambda}
 \gamma_{\mu} \slashed{F}
\tfrac 1 {16} \slashed{T}
\gamma^\mu  \ve) =   \underbrace{\mycolor{blue} \tfrac 1 {2} (\conj{\lambda} \gamma^{a} 
  \ve)
F^{\mu \nu }  T_{\mu \nu  a}  }_{\text{\textcircled{2}}}
  -\underbrace{ \mycolor{brown} \tfrac 1 {2}  (\conj{\lambda} \gamma^{\nu} \ve)
 T_{\mu \nu  a} F^{\mu a } 
 }_{\text{\textcircled{1}}}
+\underbrace{ \mycolor{purple}\tfrac 1 2  (\conj{\lambda}
\gamma^{\nu a b} 
  \ve)
 T_{\mu \nu  a} F^{\mu b } }_{\text{\textcircled{4}}}
  \end{aligned}
\end{equation}
and
\begin{equation}
\label{eq:4}
  \begin{aligned}
&  \mathbf{term_{T2}} = -\tfrac  1  4 \conj{\lambda} \slashed{D}(
 T_{\mu \nu a} \Phi^a \gamma^{\mu
  \nu} \ve ) = \\
& \qquad =  -\tfrac 1 4 (\conj{\lambda}\gamma^{\rho}
\gamma^{\mu \nu} \ve)  [D_{\rho} T_{\mu \nu a}] \Phi^a  
\underbrace{\mycolor{brown}  -\tfrac 1 2   (\conj{\lambda} \gamma^{\nu}
  \ve) T_{\mu \nu a} F_{\mu  a}  }_{\text{\textcircled{1}}} -\underbrace{ \mycolor{purple}  \tfrac 1 4  (\conj{\lambda}\gamma^{\rho \mu \nu} \ve)  T_{\mu \nu a} F_{\rho a}
}_{\text{\textcircled{4}}}
\underbrace{\mycolor{green}-\tfrac 1 {4} (\conj{\lambda} \gamma^{a} \ve) 
 T_{\mu \nu a } T_{\mu \nu b} \Phi^b }_{\text{\textcircled{3}}}
  \end{aligned}
\end{equation}
Using (\ref{eq:Tep}) all $TF$ terms cancel between (\ref{eq:4}) and
(\ref{eq:5})
and finally the $[DT]\Phi$ terms in (\ref{eq:5})(\ref{eq:6})(\ref{eq:4})
cancel as well as
\begin{equation}
  \begin{aligned}
&    (\ref{eq:5}): \qquad   (\lambda \gamma^{\nu} \ve) 
[D^{\mu} T_{\mu \nu a}] \Phi^a  \\
&    (\ref{eq:6}): \qquad  \tfrac 1 2  (D^{\mu} T_{\mu \nu a} ) \Phi_b (\conj{\lambda} \gamma^{b} 
  \gamma^{\nu a} \ve) =  - \tfrac 1 2 (\conj{\lambda} \gamma^{\nu} \ve) 
 D^{\mu} T_{\mu \nu a} \Phi^a + \tfrac 1 2 (\conj{\lambda} \gamma^{\nu
   ab} \ve) \Phi^b D^{\mu} T_{\mu \nu a} \\
&    (\ref{eq:4}): \qquad   -\tfrac 1 4 \conj{\lambda} D_{\rho} (T_{\mu \nu a}) \Phi^a \gamma^{\rho}
\gamma^{\mu \nu} \ve = - \tfrac 1 2 (\conj{\lambda} \gamma^{\mu} \ve) 
D^{\mu} T_{\mu \nu a} \Phi^a - \tfrac 1 4 (\conj{\lambda} \gamma^{\mu
  \nu \rho} \ve) \Phi^a  D_{\rho} T_{\mu \nu a} 
  \end{aligned}
\end{equation}




\newpage

\renewcommand{\refname}{References to the articles in this volume}

\renewcommand{\refname}{References}

\bibliography{libS4}

\end{document}